\documentclass[fleqn,usenatbib]{mnras}

\usepackage{newtxtext,newtxmath}

\usepackage[T1]{fontenc}

\DeclareRobustCommand{\VAN}[3]{#2}
\let\VANthebibliography\thebibliography
\def\thebibliography{\DeclareRobustCommand{\VAN}[3]{##3}\VANthebibliography}

\usepackage{graphicx}	
\usepackage{amsmath}	
\usepackage{comment}
\usepackage{placeins}

\title[The sub-Neptune GJ 3090 b observed by CRIRES+]{Limits on the atmospheric metallicity and aerosols of the sub-Neptune GJ~3090~b from high-resolution CRIRES+ spectroscopy}

\author[L. T. Parker et al.]{Luke T. Parker,$^{1}$\thanks{E-mail: luke.parker@physics.ox.ac.uk}
João M. Mendonça,$^{2,3,4}$
Hannah Diamond-Lowe,$^{2,5}$
Jayne L. Birkby,$^{1}$ \newauthor
Annabella Meech,$^{6}$
Sophia R. Vaughan,$^{1,7}$
Matteo Brogi,$^{8,9}$ 
Chloe Fisher,$^{1}$ 
Lars A. Buchhave,$^{2}$ \newauthor
Aaron Bello-Arufe,$^{10, 2}$
Laura Kreidberg,$^{7}$ 
Jason Dittmann $^{7,11}$ 
\\
\\
$^{1}$ Department of Astrophysics, University of Oxford, Denys Wilkinson Building, Keble Road, Oxford, OX1 3RH, UK\\
$^{2}$ Department of Space Research and Space Technology, Technical University of Denmark, Elektrovej 328, 2800 Kgs. Lyngby, DK\\
$^{3}$ School of Physics and Astronomy, University of Southampton, Highfield, Southampton SO17 1BJ, UK \\
$^{4}$ School of Ocean and Earth Science, University of Southampton, Southampton, SO14 3ZH, UK\\
$^{5}$ Space Telescope Science Institute, 3700 San Martin Drive, Baltimore, MD 21218, USA\\
$^{6}$ Center for Astrophysics, Harvard \& Smithsonian, 60 Garden Street, Cambridge, MA 02138, USA\\
$^{7}$ Max-Planck-Institut fur Astronomie, Konigstuhl 17, 69117 Heidelberg, Germany\\
$^{8}$ Dipartimento di Fisica, Universit\`a degli Studi di Torino, via Pietro Giuria 1, I-10125, Torino, Italy\\
$^{9}$ INAF-Osservatorio Astrofisico di Torino, Via Osservatorio 20, I-10025 Pino Torinese, Italy\\
$^{10}$ Jet Propulsion Laboratory, California Institute of Technology, Pasadena, CA 91109, USA\\
$^{11}$ Department of Astronomy, University of Florida, Gainesville, FL 32611, USA\\}

\date{Accepted 2025 March 18. Received 2025 March 10; in original form 2025 January 10}

\pubyear{2025}

\begin{document}
\label{firstpage}
\pagerange{\pageref{firstpage}--\pageref{lastpage}}
\maketitle

\begin{abstract}
The sub-Neptune planets have no solar system analogues, and their low bulk densities suggest thick atmospheres containing degenerate quantities of volatiles and H/He, surrounding cores of unknown sizes. Measurements of their atmospheric composition can help break these degeneracies, but many previous studies at low spectral resolution have largely been hindered by clouds or hazes, returning muted spectra. Here, we present the first comprehensive study of a short-period sub-Neptune using ground-based, high-resolution spectroscopy, which is sensitive to the cores of spectral lines that can extend above potential high altitude aerosol layers. We observe four CRIRES+ \textit{K}-band transits of the warm sub-Neptune GJ~3090~b (T$_{\text{eq}}$~=~693$\pm$18~K) which orbits an M2V host star. Despite the high quality data and sensitivity to CH$_4$, H$_2$O, NH$_3$, and H$_2$S, we detect no molecular species. Injection-recovery tests are consistent with two degenerate scenarios. First, GJ~3090~b may host a highly metal-enriched atmosphere with >~150~Z$_{\odot}$ and mean molecular weight >~7.1~g~mol$^{-1}$, representing a volatile dominated envelope with a H/He mass fraction $x_{\text{H/He}} < 33\%$, and an unconstrained aerosol layer. Second, the data are consistent with a high altitude cloud or haze layer at pressures <~3.3$\times$10$^{-5}$~bar, for any metallicity. GJ~3090~b joins the growing evidence to suggest that high metallicity atmospheres and high altitude aerosol layers are common within the warm (500~$<~T_{\text{eq}}~<$~800~K) sub-Neptune population. We discuss the observational challenges posed by the M-dwarf host star, and suggest observing strategies for transmission spectroscopy of challenging targets around M-dwarfs for existing and ELT instrumentation.
\end{abstract}

\begin{keywords}
planets and satellites: atmospheres -- planets and satellites: individual: GJ 3090 b -- techniques: spectroscopic
\end{keywords}



\section{Introduction}
\label{sec:Intro}

The prevalence of small exoplanets on orbits of less than 10 days is one of the most unexpected discoveries in the detection of planets beyond our solar system \citep{Batalha2014}. Population studies have revealed that these small close-in exoplanets are divided into two distinct categories: the super-Earths (R$_\text{p}$~$\lesssim$~1.7 R$_{\oplus}$) and sub-Neptunes \citep[1.7~R$_{\oplus}$~$\lesssim$~R$_\text{p}$ $\lesssim$~4~R$_{\oplus}$; e.g.][]{Rogers2015,Fulton2017,Fulton2018,VanEylen2018,Dattilo2024}. The terrestrial super-Earths have bulk densities similar to that of Earth and Venus, but sub-Neptunes have no Solar System analogues. Their low bulk densities point to thick atmospheres containing substantial quantities of hydrogen and helium alongside water and volatiles, surrounding degenerate amounts of iron and silicates in solid cores of unknown sizes \citep{Rogers2010,Dorn2017, Hu2021, Yu2021,Luque2022}. This compositional uncertainty has led to the proposition of a range of possible planetary structures including H/He dominated, low mean molecular weight atmospheres (`Gas Dwarf'; e.g. \citealt{Fortney2013, Lopez2014, Buchhave2014}); high metallicity envelopes composed of miscible H/He and metals (`Miscible Envelope sub-Neptune'; \citealt{Benneke2024}); H$_2$O dominated, high mean molecular weight atmospheres (`Water Worlds'; e.g. \citealt{Mousis2020,Aguichine2021,Pierrehumbert2023,Piaulet2023,Piaulet-Ghorayeb2024}); and H/He dominated atmospheres with a defined surface-atmosphere boundary, either a transition to a liquid water ocean (`Hycean worlds'; e.g. \citealt{Hu2021,Madhusudhan2021,Madhusudhan2023}), or a magma ocean (e.g. \citealt{Kite2020,Schlichting2022,Zilinskas2023,Shorttle2024}). These scenarios, which are degenerate when considering only the bulk planetary parameters, can be broken through atmospheric observations \citep{Bean2021}. Constraints on the mean atmospheric weight can provide boundary conditions for models of interior structure \citep{Dorn2017, Nixon2021}, while the measurement of atmospheric composition can probe surface-atmosphere exchange \citep{Misener2023,Shorttle2024}. Specifically, measurements of the atmospheric metallicity \citep{Nixon2024}, C/O ratio \citep{Seo2024}, mean molecular weight \citep{Benneke2024}, and the abundances of individual atmospheric constituents \citep{Madhusudhan2023, Shorttle2024, Yang2024}, have all been suggested to distinguish between the large range of scenarios proposed. 

H/He-rich atmospheres are predicted to produce large signals of trace species in transmission spectra due to their large scale heights, and sub-Neptunes have been targeted extensively at low resolution with HST/WFC3’s G141 grism (e.g. \citealt{Kreidberg2014,Benneke2019, Guo2020, Mikal-Evans2021,Mikal-Evans2023,Roy2023}). However, observations have predominantly returned flat or muted spectra relative to what is predicted for a clear atmosphere, suggesting high-altitude clouds or hazes (collectively aerosols) that obscure the targeted molecular features (e.g. \citealt{Kreidberg2014}). Initial population studies have suggested that these aerosol abundances, and consequently the strength of the attenuation, follow trends with equilibrium temperature, but are prevalent across a large temperature range (300~$<~T_{\text{eq}}~<$~900~K; \citealt{Morley2015,Crossfield2017,Brande2024}). The study of these enigmatic worlds at low-resolution has been transformed by the advent of precise near-infrared spectroscopy with \textit{JWST}, with molecular detections of CH$_4$, CO$_2$, H$_2$O in cool and warm sub-Neptunes (e.g. \citealt{Madhusudhan2023, Holmberg2024, Benneke2024, Piaulet-Ghorayeb2024, Davenport2025}), but the obscuration of spectral features in transmission by aerosols remains a fundamental limitation of many observations at low spectral-resolution (e.g. GJ 1214 b; \citealt{Kempton2023, Schlawin2024, Ohno2024}).

However, observations using ground-based high-resolution cross-correlation spectroscopy at R~$\approx$~100\,000 (HRCCS; see \citealt{Birkby2018}) are additionally sensitive to the cores of spectral lines that extend above the aerosol layer and are therefore predicted to be sensitive to the unique spectral lines of molecules in sub-Neptune atmospheres, even in the presence of a high altitude aerosols \citep{deKok2014,Kempton2014,Pino2018,Gandhi2020, Hood2020}. This detectability relies upon the assumption of well-mixed atmospheres, such that targeted molecules are present in sufficient abundances above the cloud deck to form strong spectral lines. Shallow atmospheres and the presence of a surface have been proposed to inhibit thermochemical kinetics and mixing on sub-Neptunes, leading to the depletion of nitrogen-based species (e.g. NH$_3$, HCN) in the presence of shallow surfaces ($\lesssim$ 10 bar; \citealt{Yu2021}), but the abundances of key molecules including CH$_4$, CO, and H$_2$O are largely invariant to this effect \citep{Hu2021, Yu2021}.

\begin{figure}
    \includegraphics[width=\columnwidth]{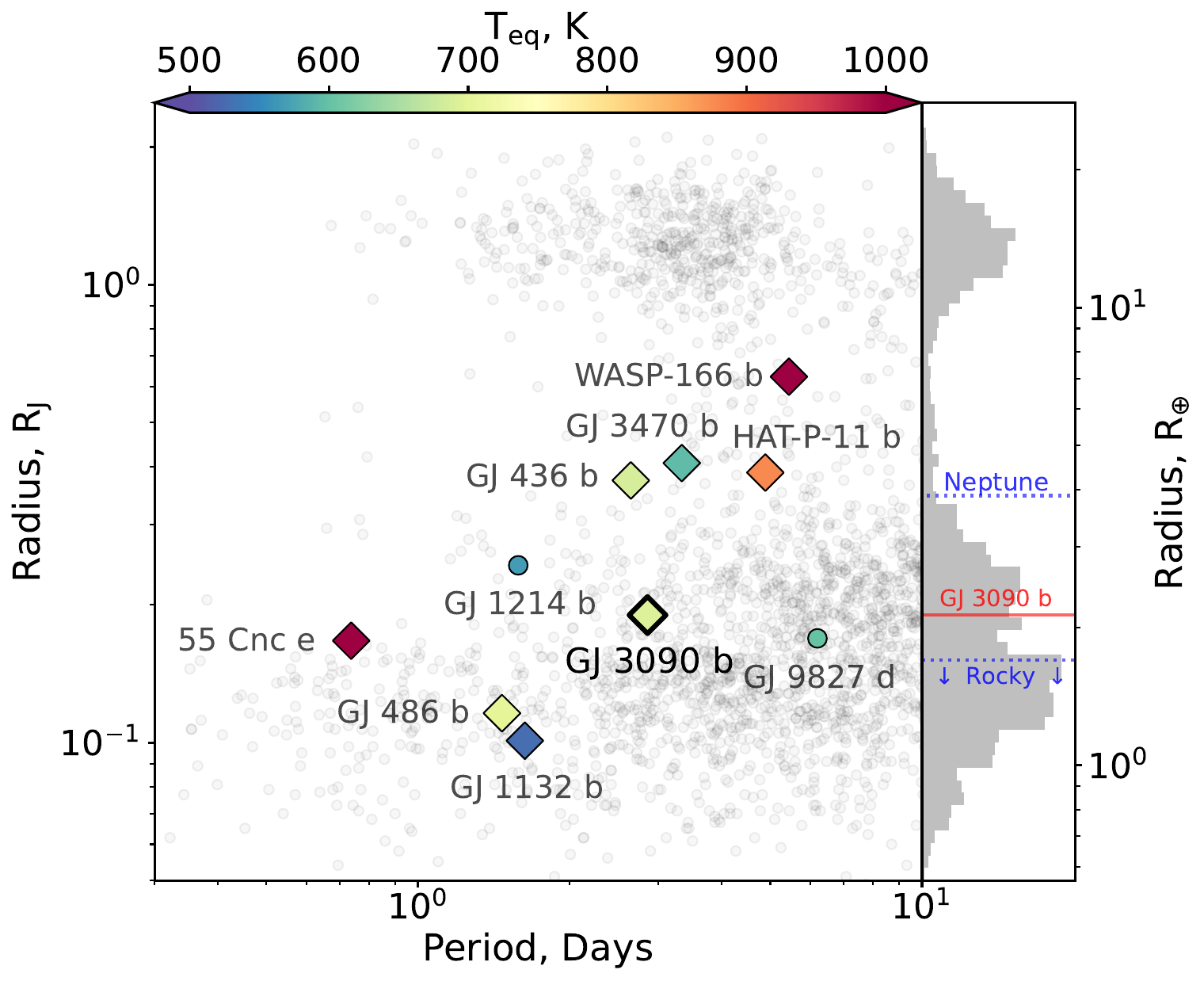}
    \caption{The sub-Neptune GJ~3090~b compared to the small population of short period, sub-Jovian planets studied with HRCCS (diamonds). Of the planets shown only HAT-P-11 b has produced confirmed molecular detections using HRCCS, while all other HRCCS studies have placed upper limits. The feasibility of HRCCS to explore the enigmatic sub-Neptune regime (with radii less than Neptune) remains largely untested. GJ~1214~b and GJ~9827~d, which have been studied in detail at low spectral resolution, are highlighted for comparison.}
    \label{fig:population_context}
\end{figure}

While transmission spectroscopy with HRCCS has proved highly successful in characterising the population of Jupiter mass planets (e.g. \citealt{Snellen2010,Brogi2016,Hoeijmakers2018_nature,Prinoth2022, Pelletier2023, Nortmann2024}), it is only beginning to push to smaller targets. In the Neptune/Saturn mass regime, \citet{Lafarga2023}, \citet{Dash2024}, and \citet{Grasser2024} observed WASP-166 b, GJ 3470 b and GJ 436 b respectively and, while they are unable to detect molecular species in the planetary atmospheres, place constraints on their metallicities and the altitude of cloud decks. \citet{Basilicata2024} report the detection of NH$_3$ and H$_2$O in the warm, Neptune-mass, HAT-P-11 b using GIANO-B, the smallest planet in which molecular detections have been found using HRCCS to date (see Figure~\ref{fig:population_context}). Attempts have also been made to observe rocky worlds with HRCCS, with efforts focused on 55~Cancri~e, none of which have detected atmospheric features \citep{Ridden-Harper2016,Esteves2017, Jindal2020, Deibert2021, Keles2022}, while some clear H/He dominated atmosphere scenarios have also been ruled out for GJ~486~b \citep{Ridden-Harper2023} and GJ~1132~b \citep{Palle2025}.

The sub-Neptune regime however, in which we anticipate a diverse range of atmospheric compositions, remains unexplored, with no reported studies of sub-Neptunes on short orbits that are readily accessible by traditional HRCCS. The only existing work on a sub-Neptune using HRCCS revolves around a single transit of TOI-732 c with IGRINS/Gemini-S \citep{Cabot2024}, attempting to extend the viability of the HRCCS technique to longer period, temperate exoplanets. The scarcity of observations is, in part, a result of the novel challenges to HRCCS observations presented by sub-Neptunes, foremost the small spectral features in transmission which necessitate bright host stars with favourable star-planet radius ratios. The most favourable sub-Neptune targets for atmospheric characterisation therefore orbit bright M-dwarf host stars, producing additional observational challenges due to their activity and spectroscopic features (see Section~\ref{sec:Obs Strat}). Recent advances in instrumentation, most notably the recent upgrade to CRIRES/VLT \citep{Dorn2014,Dorn2023}, offers the first practical opportunity to observe the atmospheres of these enigmatic sub-Neptunes at high spectral resolution. 

In this work we observe four transits of GJ~3090~b, a $M_p=3.34\pm0.72 M_{\oplus}$, $R_p=2.13\pm 0.11 R_{\oplus}$ sub-Neptune around a bright ($K=7.3$~mag) M2V dwarf \citep{Almenara2022}, see Table~\ref{tab:parameters_table}. Its low density of $\rho_p=1.89_{-0.45}^{+0.52}$~g~cm$^{-3}$ appears consistent with a primordial H/He envelope, while the location of GJ~3090~b immediately above the radius valley \citep{Fulton2017}, places it in an ideal region of the parameter space to probe the transition between super-Earths and sub-Neptunes \citep{Bean2021,Parc2024}. GJ~3090~b has an equilibrium temperature of T$_{\text{eq}}$ = 693$\pm$18 K, and is predicted to be among the most readily observable sub-Neptunes for transmission spectroscopy with a transmission spectroscopy metric (TSM; \citealt{Kempton2018}) of $221^{+66}_{-46}$, second only to GJ~1214~b \citep{Almenara2022}, under the assumption of a clear H/He dominated atmosphere. While the TSM was developed for low resolution \textit{JWST}/NIRISS observations and normalised to the J-band magnitude, it encapsulates both the predicted atmospheric scale height and the brightness of the host star in the NIR, making it a reasonable metric to also apply to \textit{K}-band HRCCS transmission observations.

With a total of 10 hours of observations, targeting the optimal spectral regions in the \textit{K}-band, we carry out the first comprehensive study of a sub-Neptune using ground-based HRCCS. We detail our CRIRES+ observations in Section~\ref{sec:Observations}. Section~\ref{sec:Methods} discusses our data processing, modelling, and analysis procedure, with results and injection tests presented in Section~\ref{sec:Results}. In Section~\ref{sec:Discussion} we discuss the constraints on the atmosphere of GJ~3090~b, and their implications for future studies, alongside recommended observing strategies for existing and future instrumentation. We conclude in Section~\ref{sec:Conclusions}. 

\begin{table}
	\centering
	\caption{The stellar and planetary parameters of the GJ~3090~b system. A22: \citet{Almenara2022}, G20: \citet{Gaia2020}, C03: \citet{Cutri2003}. \label{tab:parameters_table}}
	\begin{tabular}{ccc} 
		\hline
		Parameter & Value & Reference\\
		\hline
            \hline
             & \textbf{          GJ 3090: stellar parameters} & \\
            \hline
            RA & 01$^{\text{h}}$~21$^{\text{m}}$~45$^{\text{s}}$.390170 & G20\\
            Dec & -46\degr~42\arcmin~51.76497\arcsec & G20\\
            \textit{K} & 7.29$\pm$0.03~mag& C03 \\ 
		R$_{\star}$ & 0.516$\pm$0.016~R$_{\sun}$ & A22 \\ 
		M$_{\star}$ & 0.519$\pm$0.013~M$_{\sun}$ & A22\\ 
		T$_{\text{eff}}$ & 	3556$\pm$70~K & A22\\ 
            Spectral type & M2V & A22\\ 
            $[$Fe~/~H$]$ & -0.060$\pm$0.120 & A22 \\
            $v$sin(i) & $<$1.47$\pm$0.05~km~s$^{-1}$ & A22\\
            Distance & 22.444$\pm$0.013 pc & A22 \\
            $v_{\text{sys}}$ & 17.4095$\pm$0.0046~km~s$^{-1}$ & A22\\ 
		\hline
             & \textbf{           GJ~3090~b: planetary parameters} & \\
            \hline
            T$_{\text{eq}}$ & 693$\pm$18~K & A22\\
            R$_{\text{p}}$ & 2.13$\pm$0.11~R$_\oplus$ & A22\\ 
		M$_{\text{p}}$ & 3.34$\pm$0.72~M$_\oplus$ & A22\\
            \textit{$\rho$}$_{\text{p}}$ &  1.89$_{-0.45}^{+0.52}$ g cm$^{-3}$& A22\\
            P$_{\text{orb}}$ & 2.8531054$\pm$0.0000023 days & A22\\
            \hline
            \hline
	\end{tabular}
\end{table}

\section{Observations}
\label{sec:Observations}

We observed four transits of GJ~3090~b with the upgraded Cryogenic High-Resolution In-frared Echelle Spectrograph \citep[CRIRES+;][]{Dorn2014,Dorn2023}, mounted at the Nasmyth B focus of the VLT UT3 (Melipal). Our observations, totalling 10~h of observing time over the four transits (See Table~\ref{tab:obs_table}), were taken between 8\textsuperscript{th} Aug and 30\textsuperscript{th} Oct 2022 (Program ID: 109.232F, PI: Diamond-Lowe). Each transit had a duration $T_{14}$ = 1.281$\pm$0.024 hr \citep{Almenara2022}, and an additional out of transit baseline of $\approx T_{14}$ was observed per transit. The propagated uncertainty on the transit time of 25~s~yr$^{-1}$ leads to a transit midpoint that is constrained to a precision of $\pm$100~s, less than the duration of a single 180~s exposure with CRIRES+. The upgraded CRIRES operates as a cross-dispersed slit spectrograph and the slit width of 0.2\arcsec was chosen to achieve the maximum spectral resolving power offered by CRIRES+, nominally R~$\sim$~92~000 when the resolution is limited by the slit width. The Multi-Application Curvature Adaptive Optics system (MACAO; \citealt{Paufique2004}) system is used with GJ~3090 as the AO natural guide star. The use of MACAO provides both an enhanced S/N and, in good observing conditions, a spectral resolution R~$\gg$~92~000, limited by the PSF width. We observed in the K2166 grating setting, which provides discontinuous wavelength coverage across the range 1.92~-~2.47~$\mu$m over seven orders. The K-band provides sensitivity to key atmospheric species predicted to be found in warm sub-Neptunes including H$_2$O, CH$_4$, CO, CO$_2$, NH$_3$, and H$_2$S. Furthermore, the M-dwarf host is bright at K-band wavelengths ($K=7.3$~mag), and the spectra remain relatively unaffected by thermal background noise compared to observations at longer wavelengths (e.g. \citealt{Parker2024}). The K2166 grating is specifically chosen as it provides a good coverage of target spectral features and has a shallow blaze function, ensuring high S/N spectra across each spectral order. Exposures were taken in the classical ABBA nod pattern (NDIT=1, NEXP=1), during which the telescope was nodded $6^{\prime\prime}$ along the slit, to allow accurate background subtraction. Transits were selected for the optimum airmass and achieve excellent observing conditions, with an average airmass of 1.15 and an average DIMM measured seeing of 0.53\arcsec across the four transits.

\subsection{Archival data: two CRIRES+ K2148 transits}

\begin{figure}
    \includegraphics[width=\columnwidth]{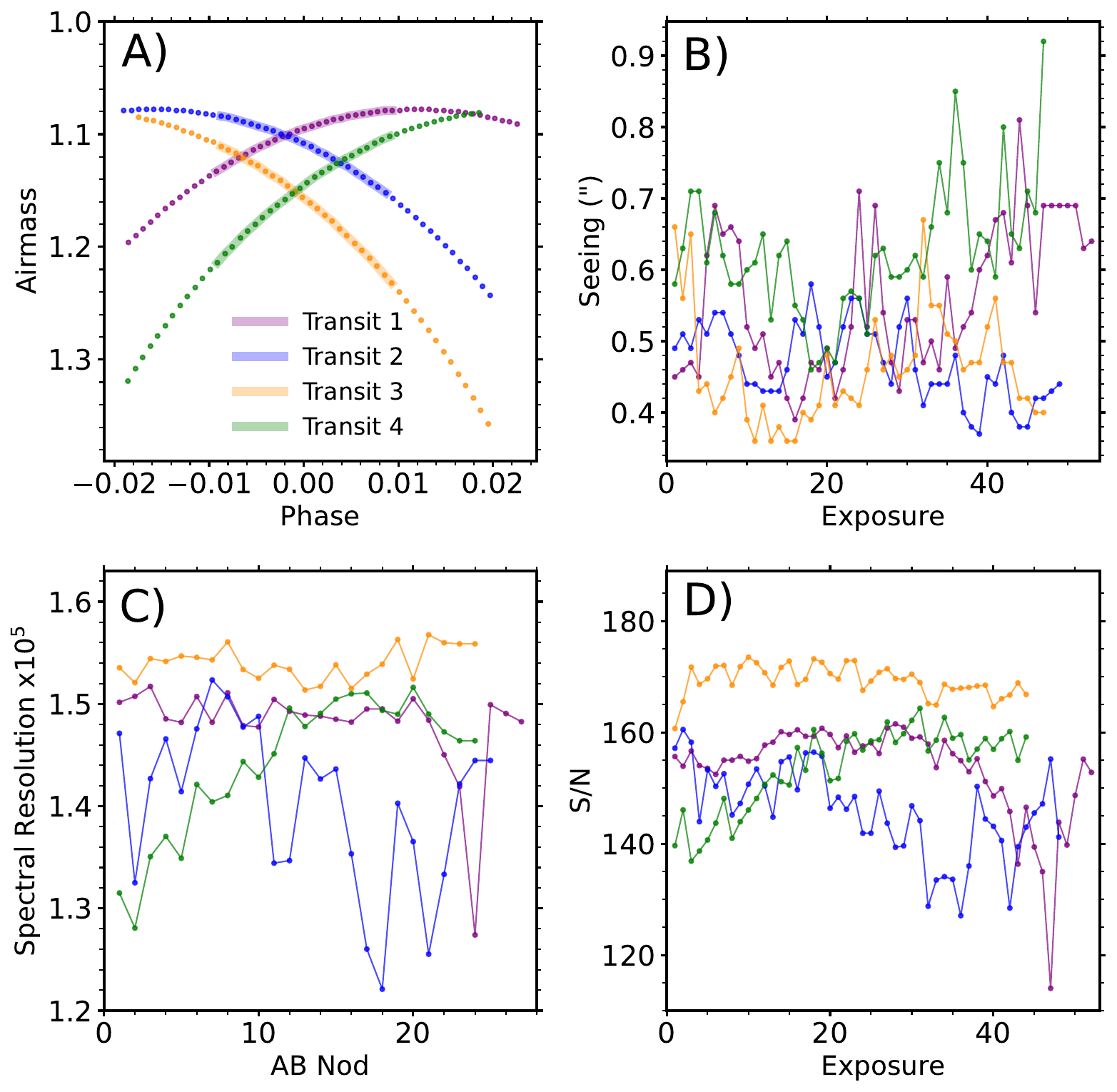}
    \caption{The observing conditions for the four CRIRES+ K2166 transits analysed in this work. A) airmass of each observed transit, optimised to occur near zenith. B) seeing conditions over the course of each transit, as measured by DIMM sensors at the observatory, which remain largely stable per night. C) measured spectral resolution of the exposures in each AB nodding pair. Due to the excellent observing conditions and MACAO performance, this is controlled by the PSF width, and is thus far in excess of the nominal slit width limited resolution of CRIRES+, R~$\sim$~92\,000. D) average S/N per spectral bin for each exposure.}
    \label{fig:airmass}
\end{figure}

We additionally analyse two archival CRIRES+ transits of GJ~3090~b taken using the K2148 grating. These transits total an additional 5.5 hours of observing time, observed on the nights of 11\textsuperscript{th} Aug and 3\textsuperscript{rd} Sept 2023 (Program ID: 0111.C-0106 GTO, PI: Nortmann). The CRIRES+ slit width of 0.2\arcsec and MACAO were used and exposures were taken in the classical ABBA nod pattern (NDIT=1, NEXP=1). The use of the K2148 grating setting,  which partially overlaps with the K2166 setting used in the primary dataset, provides discontinuous wavelength coverage across the range 1.97~--~2.45~$\mu$m over six usable orders. However, in comparison to the excellent observing conditions achieved over the four K2166 transits which form the primary dataset in this work (see Figure~\ref{fig:airmass}), the additional transits are taken in more challenging conditions. The average seeing was highly variable across both transits, with an average of 0.81\arcsec and 1.10\arcsec for the two K2148 transits respectively, while the radiometer measured precipitable water vapour (PWV) for both transits is variable over the night, reaching maxima of 2.8 and 4.31 ppm. The inclusion of these additional transits therefore do not appreciably improve the strength of our constraints on the atmosphere of GJ~3090~b. Nonetheless, we examine the limitations of these data in comparison to the highly constraining K2166 transits, and discuss the lessons learned for future observations of sub-Neptunes with HRCCS in Section~\ref{sec:archival_data} and Section~\ref{sec:Obs Strat}.

\begin{table}
	\centering
	\caption{Details of the 1.92--2.47~$\mu$m K2166 CRIRES+ observations of GJ~3090~b. \label{tab:obs_table}}
	\begin{tabular}{lcccc}
		\hline
             & Transit 1 & Transit 2 & Transit 3 & Transit 4\\
		\hline
            \hline
            UTC Date & 08/08/2022 & 07/10/2022 & 27/10/2022 & 30/10/2022\\
            Avg.~PWV & 2.26~mm & 1.65~mm & 0.38~mm & 0.64~mm \\
            Avg.~Seeing & 0.56\arcsec & 0.47\arcsec & 0.46\arcsec & 0.62\arcsec\\
            Avg.~Airmass & 1.11 & 1.12 & 1.18 & 1.17 \\
            N$_{\text{exp}}$ & 54 & 50 & 48 & 46\\
            DIT & 180~s & 180~s & 180~s & 180~s\\
            Slit Width & 0.2\arcsec & 0.2\arcsec & 0.2\arcsec & 0.2\arcsec\\
            Resolution & 148\,000 & 140\,000 & 153\,000 & 144\,000 \\
            $v_{\text{bary}}$ at T$_0$ & 11.68~km/s & -6.92~km/s & -12.72~km/s & -13.19~km/s \\
		\hline
            \hline
	\end{tabular}
\end{table}

\section{Methods}
\label{sec:Methods}

\begin{figure}
    \includegraphics[width=\columnwidth]{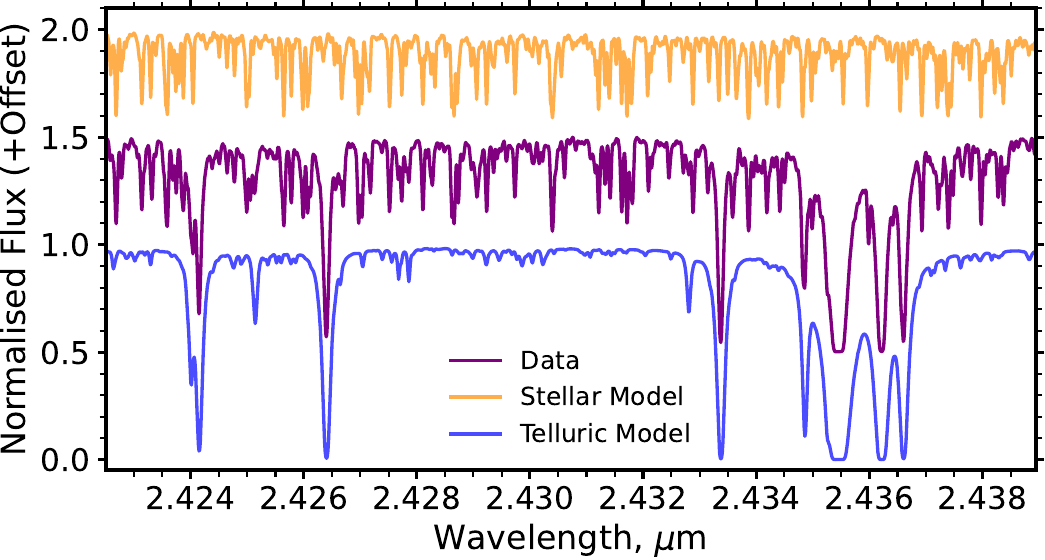}
    \caption{The stellar (orange) and telluric (blue) contributions to the observed CRIRES+ spectra of GJ 3090 (purple). Note the near universal wavelength coverage of stellar lines, rendering established analysis techniques such as masking stellar lines unfeasible, and the simultaneous modelling of telluric and stellar spectra highly challenging.}
    \label{fig:model_vs_data}
\end{figure}

\subsection{Basic calibrations}
\label{sec:basic_cals}

Basic calibrations are carried out using \textsc{pycrires}\footnote[1]{\url{https://pypi.org/project/pycrires/}} \citep{pycrires,Landman2024}, an open source python wrapper for the EsoRex cr2res routines\footnote[2]{\url{https://www.eso.org/sci/software/pipe_aem_table.html}}. The raw frames are flat-fielded and dark subtracted to remove detector and readout artefacts present in the raw exposures. Detector non-linearity corrections are applied, bad pixels are flagged and the imprinted blaze function is corrected. Each exposure (A or B) is subsequently subtracted from its corresponding nodding pair (B or A), providing an effective background subtraction as the spectral trace at the A and B nodding positions fall on different regions of the detector. From each background-subtracted exposure a 1D science spectrum of the star is optimally extracted at both the A and B nod positions for each spectral order \citep{Horne1986}, considering the slit-image curvature, which in the \textit{K}-band is accurately calibrated using the Fabry-Pérot Etalon (FPE) System \citep{Dorn2023}. In the K2166 grating setting, seven spectral orders are dispersed across the CRIRES+ detector focal plane array, consisting of three adjacent Hawaii 2RG detectors, each with dimensions of $2048\times2048$ pixels. Further details of the initial spectral extraction can be found in Appendix \ref{app:CRIRES+_reduction}. Due to the excellent performance of the CRIRES+ MACAO system, the stellar point spread function (PSF) is sufficiently narrow that it does not evenly fill the 0.2$\arcsec$ slit. The PSF is measured to have an average full-width half-maximum (FWHM) of $\approx$~2.2 pixels in the spatial direction of the CRIRES+ detectors, where a FWHM of 3.6 pixels would correspond to the width of the 0.2$\arcsec$ slit. This effect, known as super-resolution, produces observations with a spectral resolution determined by the PSF width, rather than the slit width, resulting in a much higher than average resolving power of R~=~146\,000 across the four transits. The resolving power is measured from the observed PSF width in the spatial direction following \citet{Nortmann2024}.

Due to the narrow AO-assisted PSF, the PSF centre can shift within the slit over the course of the observations. If we assume that the motion of the PSF is stochastic, we can use the vertical shift of the PSF within the slit as a proxy for the shift across the slit width. We measure an average standard deviation in the vertical placement of the PSF of 0.00672$\arcsec$, which would correspond to shifts of 3$\%$ of the slit width. Despite the stability of the PSF core in the slit this can result in shifts on the detector on the order of a few pixels between spectra extracted from the A and B nodding positions for each transit. This effect is independent of the offset of spectra from A and B nods which arises from the misalignment of the CRIRES+ slit tilt with respect to the detector columns, which is accounted for in the wavelength calibration. Furthermore, the narrow FWHM of the stellar signal results in enhanced contamination from individual bad pixels that intersect the trace at either nod position, leading to discrepant systematic effects at the pixel level for the spectra extracted from the A or B nod positions. In order to limit the impact of these offsets and nod-dependent systematic effects, spectra at the A and B nod positions are subsequently reduced independently, and only combined following cross-correlation.

All spectra from each nodding position are aligned to a common wavelength grid for the nod and ordered in phase, with the grids of both the A and B spectra maintaining the original detector pixel scale. Initial wavelength calibration of the spectra is carried out using the UNe lamp as the stable wavelength reference source, with additional corrections applied by the FPE. Finally, to ensure an accurate wavelength calibration a telluric model from $\textsc{molecfit}$ \citep{Smette2015} is used as the stable reference source, and cross-correlated with the second spectrum of each night to produce a refined wavelength solution for each nodding position.

An alternative solution to the CRIRES+ AB nod offsets is to align all the extracted spectra from both nodding positions onto a common wavelength grid (e.g. \citealt{Nortmann2024}). We additionally test this method, aligning the A and B nods from each nodding pair by cross-correlating each AB nod pair with each other, then aligning to a common wavelength grid. While this approach produced excellent alignment between the spectra, the combination of nod-specific systematics from the A and B nods adversely affected the cleaning process, resulting in a marginally lower detection significance when carrying out injection recovery tests. As this method requires additional interpolation of the primary data products, we therefore reduce the extracted spectra from the A and B nod positions independently.

\begin{figure}
    \includegraphics[width=\columnwidth]{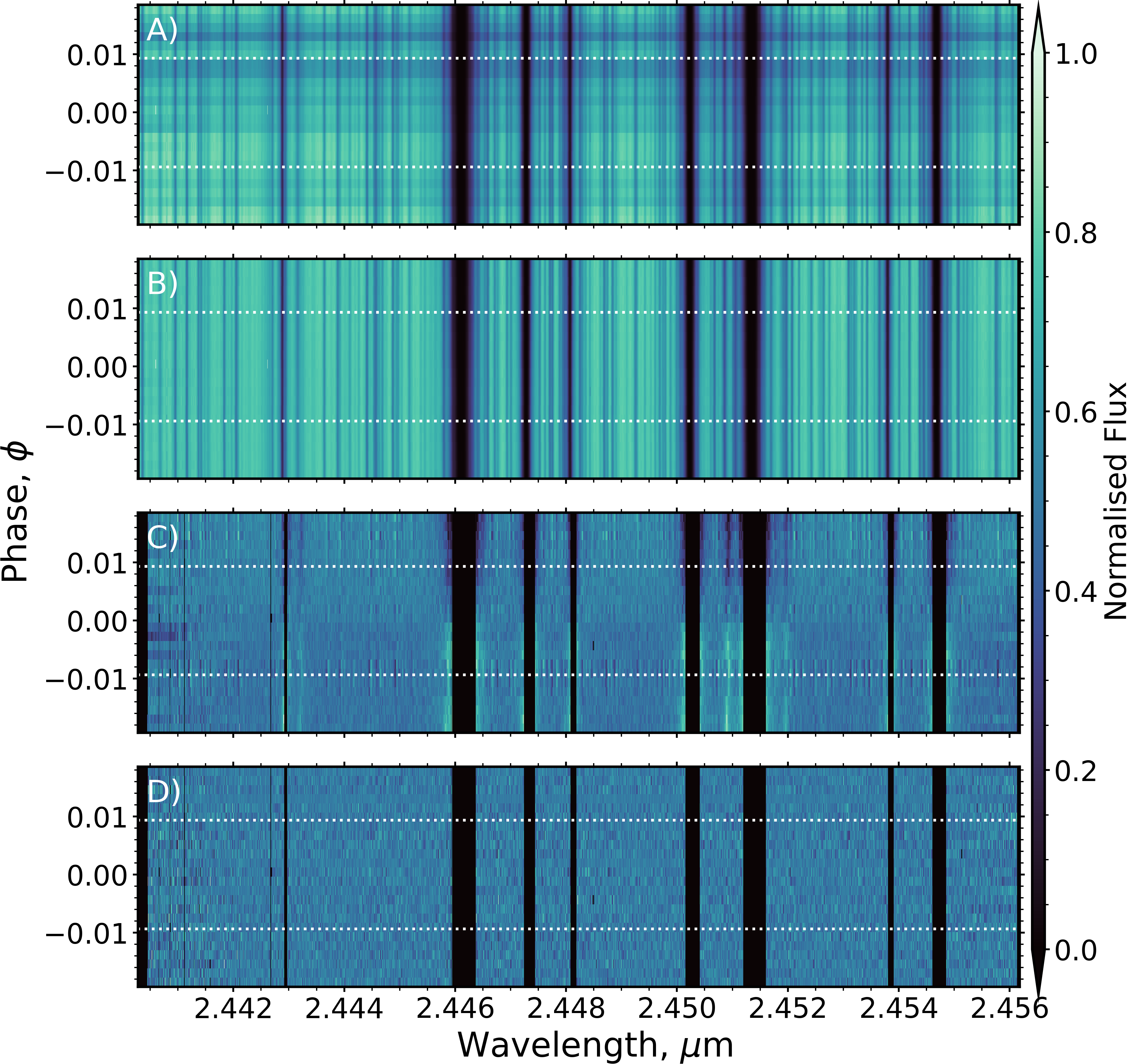}
    \caption{The data processing stages for the four CRIRES+ transits, with transit ingress and egress marked by the white dashed horizontal lines. Panel A shows the optimally extracted and phase ordered spectra, following basic calibrations including background subtraction. Panel B shows the normalised, wavelength calibrated spectra, where any residual large scale trends have been removed through division by a smoothed spectrum. Panel C shows the spectra shifted into the stellar rest frame, where saturated telluric lines and high variance columns have been masked following bad pixel correction and division by the master spectrum constructed from the mean of all exposures. Finally, Panel D shows the spectra after \textsc{sysrem} is applied iteratively to remove remaining linear systematic trends in the data. These resulting cleaned spectra are subsequently cross-correlated with planetary models. This figure shows the B nod spectra extracted from the first order of the second detector during the second transit.}
    \label{fig:data_proc}
\end{figure}

\subsection{Post-Processing}
\label{sec:post_proc}
At this stage in the reduction process the dominant features in the spectra are telluric absorption features and stellar lines from the M-Dwarf host star GJ 3090 (see Figure~\ref{fig:model_vs_data}). Each of these components is stationary with time within its own reference frame, and we choose to perform our cleaning procedure in the stellar rest frame, which successfully removes contamination from the M-dwarf host, without significantly compromising the removal of telluric features \citep{Nortmann2024}. The stellar features are ubiquitous across the wavelength range observed, and residuals from their imperfect correction have the largest impact in cross-correlation space (see Section \ref{sec:star_and_telluric_modelling}), as they overlap with the systemic velocity of potential planetary signals, and we therefore prioritise their correction when removing systematics from the data.

Our cleaning procedure is as follows (see Figure~\ref{fig:data_proc}). First, the extracted spectra are normalised and columns with $< 20\%$ transmission are masked to remove the cores of saturated telluric lines, while 15 columns at both edges of each order which show enhanced systematics are additionally masked. Any remaining outliers on the scale of a single pixel are detected through application of a Laplacian of the Gaussian algorithm, or `Blob Detection Algorithm' (e.g. \citealt{Kong2013}, see \citealt{vanSluijs2023}), and pixels that deviate by $>$~5$\sigma$ are masked. Following this, the spectra are shifted into the stellar rest frame through flux conserving interpolation. For each night a master spectrum, containing stellar and telluric spectral features, is created using the mean of all spectra for each nodding position. Due to the low number of frames available per nodding position ($\sim$25) and the 1:1 ratio of T$_{\text{in}}$:T$_{\text{out}}$ frames, we construct this master spectrum using all frames, not just the out-of-transit frames. Subsequently, each spectrum is divided by the master spectrum of the corresponding nodding position (Panel C; Figure~\ref{fig:data_proc}). As the planet spectral lines are Doppler-shifted by $\sim$14~km~s$^{-1}$ across the in-transit frames, the planetary signal is rapidly shifting between spectral pixels. Any self-subtraction of the planetary signal through the inclusion of the in-transit frames is therefore $\lesssim 4\%$. Injection-recovery tests confirm that the higher precision master spectrum produced when using spectra from all exposures sufficiently negates any possible self-subtraction of the planet. Following the division by the master spectra columns with standard deviations $>$~4$\sigma$ are masked to remove high variance columns.

At this stage significant residuals still remain, which are removed through the application of the de-trending algorithm \textsc{sysrem} \citep{Tamuz2005,Mazeh2007,Birkby2013}. \textsc{sysrem} iteratively identifies and removes linear systematic trends from the data, incorporating the error of each datapoint. However, \textsc{sysrem} requires the adoption of a robust stopping criteria to prevent erosion of the planetary signal, and to avoid the optimisation of systematic noise. Here, we apply the data-driven approach of Spring $\&$ Birkby (in preparation), and calculate the $\Delta \varsigma$ metric for each iteration, defined as:

\begin{equation}
    \Delta \varsigma  = \frac{^{(i-1)}\sigma - ^{(i)}\sigma}{^{(i-1)}\sigma} 
\end{equation}

where $^{(i-1)}\sigma$ and $^{(i)}\sigma$ are the standard deviation of the data before and after the $i$-th iteration of \textsc{sysrem}, respectively. This metric therefore encapsulates the percentage change in the standard deviation of the data for the application of each \textsc{sysrem} iteration. The stopping point of \textsc{sysrem} is selected to be the iteration at which $\Delta \varsigma$ plateaus, and is calculated independently for each detector order (See Figure \ref{fig:sysrem_it}). The number of \textsc{sysrem} components removed for each data set is displayed in Table \ref{tab:sysrem_table}.

\subsection{Atmospheric models}
\label{sec:atmo models}

\subsubsection{Atmospheric models for cross-correlation}
\label{sec:CCF models}
\begin{figure*}
    \includegraphics[width=2.1\columnwidth]{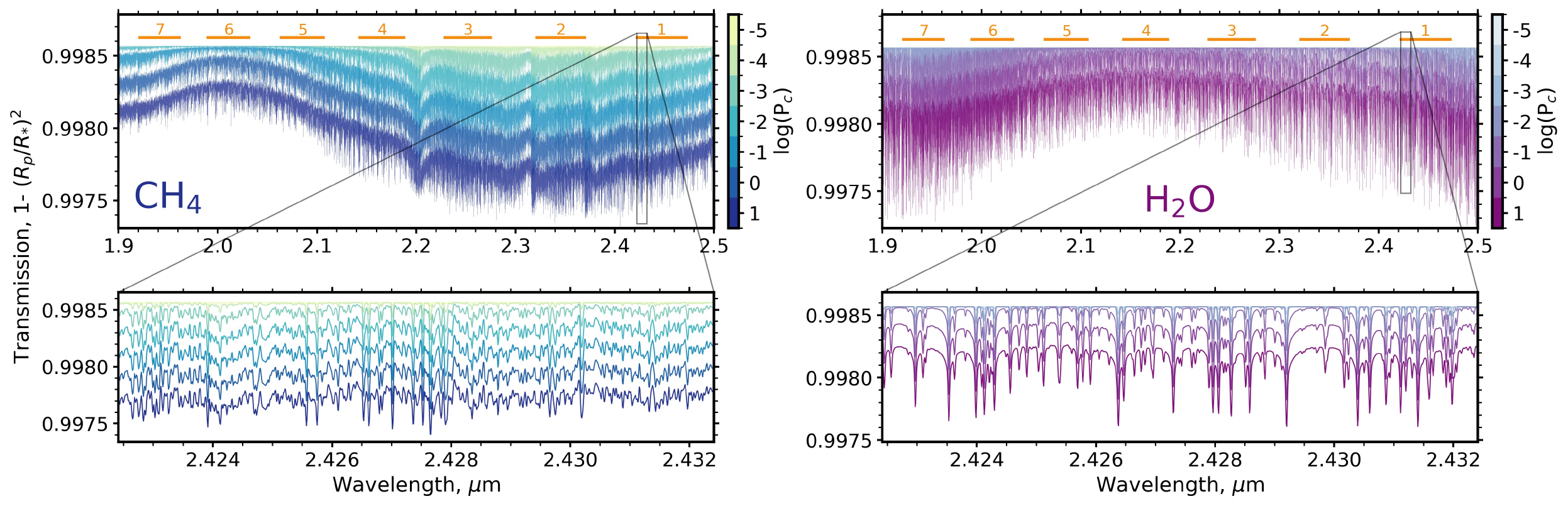}
    \caption{High spectral resolution models of GJ~3090~b with CH$_4$ (left) and H$_2$O (right) as the only spectroscopically active atmospheric constituents. The colour gradient delineates models with an aerosol layer at pressure level P$_{\text{c}}$ in bar. A rising aerosol layer truncates spectral lines in the transmission spectrum, reducing their detectability, although this effect is less pronounced than at lower spectral resolution as the cores of the R = 146\,000 lines extend above the aerosol layer. For each model, P$_0$ is fixed to the cloud deck pressure. The wavelength regions covered by the CRIRES+ K2166 grating are highlighted in orange.}
    \label{fig:cloud_models}
\end{figure*}

In order to extract the faint planet spectrum from the residual noise in the processed data, we need to combine the signal from each of the planetary spectral lines. We achieve this via cross-correlation with high-resolution model spectra derived from atmospheric models (Figure~\ref{fig:cloud_models}). To generate the spectral templates, we calculate how the apparent size of the planet changes for a grid of of atmospheric gases and cloud top levels.

For our transmission model templates, we have taken into account the contribution from the dominant and most readily detectable molecular species from the following line list databases: CO$_2$ \citep{Rothman2010}, CO \citep{Li2015}, CH$_4$ \citep{Hargreaves2020}, H$_2$O \citep{Polyansky2018}, H$_2$S \citep{Azzam2016}, and NH$_3$ \citep{Coles2019}. The absorption cross sections for each molecule at different pressures and temperatures were calculated using the open-source custom opacity calculator \texttt{HELIOS-K} \citep{Grimm2021}, where we assume Voigt line profiles for the absorption lines, 0.01~cm$^{-1}$ spectral resolution, and a constant line cutoff of 100~cm$^{-1}$. We have also taken into account the Rayleigh-scattering contribution from each molecular species and the collision-induced absorption (CIA) from H$_2$-H$_2$ and H$_2$-He \citep{Richard2012}. The concentrations for each chemical species were calculated using the \texttt{FastChem} model \citep{Stock2018}. The atmospheric composition structure of GJ~3090~b is still poorly constrained observationally, and its values may depart from chemical equilibrium by processes such as atmospheric dynamics, condensation chemistry, and photochemistry (e.g. \citealt{Venot2014,Mendonca2018b}). However, due to the poor constraints on these processes, we adopt a simple prescription for the atmosphere probed by the transmission spectra, assuming chemical equilibrium, and an isothermal profile at the equilibrium temperature (a reasonable approximation given the S/N of the planetary spectrum, e.g. \citealt{Young2024}). Cloud decks in our models are represented by a layer of large particles that act as grey absorbers, providing a parametrisation that is agnostic to the specific aerosol (cloud or haze) species that results in obscuration of the spectra. Our model calculates the transmission spectra with R~$\sim$~250\,000 following the formalism presented in \cite{Gaidos2017}, \cite{Bower2019} and \cite{Bello-Arufe2023}, and the spectra are subsequently broadened to the measured instrumental resolution (R~$\sim$~146\,000) prior to cross-correlation. The model computes the effective tangent height in an atmosphere discretised into 200 annuli. The code operates on GPUs and effectively generates a grid of atmospheric models for various metallicity levels and cloud tops used in this work. We note that, while GJ 3090 b has precise measurements of the planetary mass and radius \citep{Almenara2022}, the uncertainty on these parameters propagates into an uncertainty on the planetary scale height of 33$\%$.

\subsubsection{Self-consistent atmospheric models}
\label{sec:self-consitent models}

In Section~\ref{sec:Discussion}, we compare the temperature profiles of two different scenarios in the atmosphere of GJ~3090~b to the saturation vapour pressure profiles of various cloud compositions. This is to evaluate the possible species that might condense in its atmosphere, and these models are not used for cross-correlation. The temperature profiles are calculated using the \texttt{HELIOS} model \citep{Malik2017,Malik2019}, and the gas concentrations are computed by the \texttt{FastChem} model \citep{Stock2018}, under the assumption of atmospheric chemical equilibrium. The \texttt{HELIOS} model employs k-distribution tables for opacities, integrating radiative fluxes over 383 spectral bands and 20 Gaussian points. Opacities are calculated using the \texttt{HELIOS-K} model \citep{Grimm2021}. We have included the main absorption species: H$_2$O \citep{Barber2006}, CO$_2$ \citep{Rothman2010}, CO \citep{Li2015}, CH$_4$ \citep{Yurchenko2014}, NH$_3$ \citep{Yurchenko2011}, HCN \citep{Harris2006}, PH$_3$ \citep{Sousa-Silva2015}, C$_2$H$_2$ \citep{Gordon2017}, H$_2$S \citep{Azzam2016}, Na and K \citep{Burrows2000,2011Kurucz}. Additionally, we incorporate Rayleigh scattering cross-sections for H$_2$ and He \citep{Lee2004,Sneep2005}, and the CIA from H$_2$-H$_2$ and H$_2$-He \citep{Richard2012}.

The incoming stellar radiation in the model is represented by an interpolated \textsc{phoenix} model \citep{Husser2013}. \texttt{HELIOS} represents the atmospheric convection by mixing the enthalpy instantaneously in a buoyant unstable atmospheric region \citep{Mendonca2018,Malik2019}, and the heat redistribution was set to 0.5, the default value for Sub-Neptune simulations with \texttt{HELIOS} \citep{Malik2017}. The model is integrated until the radiative-convective equilibrium criterium is achieved in every layer of the model.

\subsection{Searching for planetary signals through cross-correlation}
\label{sec:cross-correlation}

Following the removal of the stellar signal and telluric contamination, the residual spectra are primarily composed of noise, but additionally contain a contribution from the planetary spectrum, buried in the noise (Panel D; Figure~\ref{fig:data_proc}). To search for and extract this weak signal (S/N~<~1 per line) we cross-correlate with model atmospheric spectra (see Section~\ref{sec:atmo models}) through calculation of the Pearson correlation coefficient, $\rho$:
\begin{equation}
\rho_{X,Y} = \frac{C_{X,Y}}{\sqrt{C_{X,X} \cdot C_{Y,Y}}}
\end{equation}

where $\rho_{X,Y}$ is the Pearson correlation coefficient between two matrices $X$ and $Y$, and $C$ is the covariance. A cross-correlation function (CCF) is produced for each model by calculating the correlation coefficient across systemic velocity shifts of $\pm$300 km~s$^{-1}$, sampled at the CRIRES+ pixel velocity resolution of 1.5~km~s$^{-1}$, and across orbital semi-amplitudes of $\pm$300 km~s$^{-1}$. The planetary signal Doppler shifts by a velocity:

\begin{equation}
v_{\text{shift}} = v_{\text{bary}} + v_{\text{sys}} + K_{\text{p}} \text{sin}( 2 \pi \phi_{\text{p}})
\end{equation}

where $v_{\text{bary}}$ is the Earth's barycentric velocity offset, $v_{\text{sys}}$ the systemic velocity of the host star relative to the Solar barycentre, $K_{\text{p}}$ the semi-amplitude of the planetary orbit, and $\phi_{\text{p}}$ the orbital phase at the time of observation. For GJ~3090~b the systemic velocity is $v_{\text{sys}}$ = 17.4095$\pm$0.0046~km~s$^{-1}$ \citep{Almenara2022}, and the orbital semi-amplitude is calculated to be $K_{\text{p}}$~=~121$\pm$1~km~s$^{-1}$ under the assumption of a circular orbit. We subsequently construct $K_{\text{p}}$--$v_{\text{sys}}$ maps for each transit and nodding position, alongside the map for the combined dataset. When searching for molecular signals from the planetary atmosphere the individual CCFs from each order are combined using a weighted sum, in which we weight each map by the average observed S/N of the spectra used to construct it, and by the number of unmasked data points used in the calculation of the CCF. For each molecule we additionally exclude spectral orders containing no spectral features of the target molecule to prevent the additional inclusion of noise in the CCF. These maps are then used to search for molecular signatures in the planetary spectrum (Section~\ref{sec:Results}), in which a signal is considered a detection if it is located at the specific systemic velocity ($v_{\text{sys}}$) and orbital semi-amplitude ($K_{\text{p}}$) of GJ~3090~b, and has a S/N~>~5. The S/N is calculated by dividing each row in the $K_{\rm p}-V_{\rm sys}$ matrix by the standard deviation of that row. Systemic velocities within $\pm20$ km~s$^{-1}$ of the planetary systemic velocity are excluded from the calculation of standard deviation to avoid the inclusion of potential planetary signals in the noise estimate.

\subsection{Log-likelihood mapping for injection tests} 
\label{sec:model_reproc}

The Pearson cross-correlation coefficient and the S/N detection metric are well suited for searching for planetary spectral features, due to the sensitivity to the location and depth of the spectral lines. However, in order to place statistically robust upper limits on the planet properties through injection tests (See Section~\ref{sec:Inj Tests}), a more strict metric is required to determine the significance at which an injected model can be detected. We therefore adopt CCF to Log-likelihood mapping, developed for Bayesian parameter estimation \citep{Brogi2019}, which enables a statistical comparison between the respective fits of successive injected models. When calculating Likelihood values, we must now account for the fact that the planetary spectral lines in the data are modified by the cleaning process used \citep{Brogi2019}. Both the division of the master spectrum and the application on detrending algorithms (in this case \textsc{sysrem}) modify and distort the relative depths of the planetary spectral lines \citep{Gibson2022,Gandhi2022}. We must therefore apply the same line distortions to the model templates used to cross-correlate with the data, in a process known as model reprocessing \citep{Dash2024}. The process adopted is as follows. When detrending the data, we save the components removed by the division of the master spectrum and by each \textsc{sysrem} iteration. Prior to each cross-correlation, we construct a noiseless matrix of the model to be cross-correlated, incorporating the specific systemic ($v_{\text{sys}}$) and planetary ($K_{\text{p}}$) velocities. This matrix is subsequently reprocessed through application of the saved components, such that the model template has undergone the same processing as the planetary spectra in the data. Following the reprocessing of the model templates, we calculate the log likelihood, log(L), following \citet{Brogi2019}:

\begin{equation}
\text{log}(L) = -\frac{N}{2} \text{log}[s_f^2 -2R(s) + s_g^2]
\end{equation}

in which $s_f^2$ is the variance of the data, $s_g^2$ the variance of the model, and $R(S)$ the cross-covariance

\begin{equation}
s_f^2 = \frac{1}{N} \sum_{n} f^2(n)
\end{equation}

\begin{equation}
s_g^2 = \frac{1}{N} \sum_{n} g^2(n-s)
\end{equation}

\begin{equation}
R(S) = \frac{1}{N} \sum_{n} f(n)g(n-s)
\end{equation}

which maps to the cross-correlation coefficient by 

\begin{equation}
C(S) = \frac{R(s)}{\sqrt{s^2_f s^2_g}}
\end{equation}

equivalent to the Pearson coefficient $\rho$. Here the value $f(n)$ represents the mean-subtracted values of a row from each order of the processed data, and $g(n)$ the corresponding processed model template, where $n$ is an individual spectral channel, $s$ denotes a wavelength shift, and $N$ is the total number of spectral pixels used in the calculation. We calculate log(L) values for each transit, order, and detector individually and these are co-added without additional weighting for each injection test, as the log(L) calculation accounts for the weighting of each section of the data. Following the calculation of the log(L) for each $K_{\text{p}}$--$v_{\text{sys}}$ matrix, these are converted to confidence intervals using a likelihood ratio test \citep{Pino2020}. Following \citet{Lafarga2023} and \citet{Dash2024}, the Likelihood ratio statistic ($\lambda$) is defined as 

\begin{equation}
\lambda = 2 [\text{log}(L_{\text{max}}) - \text{log}(L)]    
\end{equation}

in which $\text{log}(L_{\text{max}})$ is the maximum likelihood within the $K_{\text{p}}$--$v_{\text{sys}}$ matrix, corresponding to the peak of the planet signal in the case of a strong detection. Following Wilks' theorem \citep{Wilks1938}, the Likelihood ratio statistic follows a $\chi^2$ distribution. As we explore two parameters, the systemic velocity of the host star ($v_{\text{sys}}$) and the semi-amplitude of the planetary orbit ($K_{\text{p}}$), this distribution has two degrees of freedom for each planetary model tested. The $p$-value of this $\chi^2$ distribution is subsequently calculated, from which the confidence intervals are derived in units of standard deviation ($\sigma$). The model with the highest log(L), corresponding to the best fit to the data, therefore has $\sigma$ = 0, and models with increasingly poor fits to the data have increasing $\sigma$.

\subsubsection{Metric for calculating the confidence at which an injected model can be excluded}
In the case of non-detection, when seeking to place upper limits on the planetary properties, we require an additional metric to rule out groups of atmospheric models. To transform the $\lambda$ metric into the confidence at which a model is excluded in an injection-recovery test (see Section~\ref{sec:Inj Tests}), we subtract the confidence value at the planetary position (denoting the best fitting model in the case of a strong detection) from the mean confidence value of the matrix which encapsulates the `noise' in the confidence map, excluding the $\pm20$ km~s$^{-1}$ $v_{\text{sys}}$ range surrounding the planetary systemic velocity. This $\Delta\sigma$ metric is thus a reformulation, in log(L), of the S/N metric commonly used with cross-correlation values in HRCCS (which simply compares the peak of the CCF for the detected planet signal to its standard deviation e.g. \citealt{Brogi2014}). Note that this metric differs from the log(L) metric used in injection- recovery grids shown in previous works as it provides an absolute detection significance for each recovered model, in comparison to the metric used in \citet{Lafarga2023} and \citet{Dash2024} which measures the quality of fit for each recovered model with comparison to the best fitting recovered model.

\section{Results}
\label{sec:Results}

\begin{figure}
    \includegraphics[width=\columnwidth]{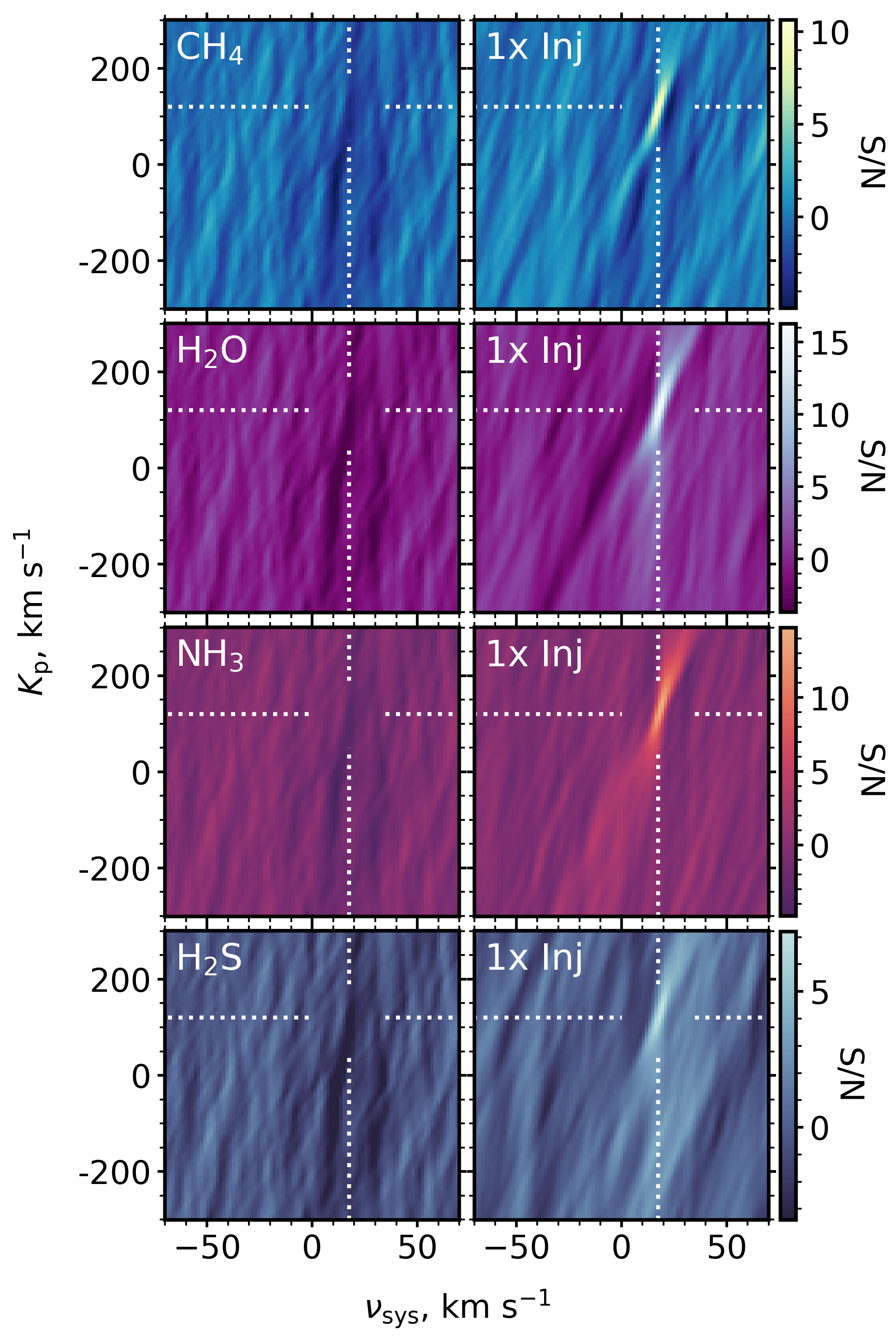}
    \caption{Left: the combined $K_{\text{p}}$--$v_{\text{sys}}$ map from the four K2166 transits of GJ~3090~b for cross-correlation with CH$_4$, H$_2$O, NH$_3$ and H$_2$S templates (1 bar cloud deck, 10~Z$_{\odot}$ metallicity). No signals are observed at the expected planetary orbital and systemic velocities at sufficient significance to support a planetary origin. Note the anti-correlation at the planetary $K_{\text{p}}$, which is offset by $\sim$5 km~s$^{-1}$ from the planetary systemic velocity. This likely arises from the overlap of stellar and telluric residual noise when combining the four transits, but does not impact the recovery significance of injected models when using the log(L) mapping. Right: the recovery of the same models (1 bar cloud deck, 10~Z$_{\odot}$ metallicity) injected into the 4 transit datasets. Each of these models can be recovered at high S/N, demonstrating the constraining power of the data.}
    \label{fig:real_vs_inj}
\end{figure}

We observe no robust detections (S/N~$>5$) of molecular species in the atmosphere of GJ~3090~b when using both the primary dataset of four K2166 transits (10 hr, excellent data quality), and upon the addition of the two K2148 transits (5.5 hr, average-poor data quality). Principal challenges include the stellar and telluric contamination which reduce the achievable contrast limits on planetary spectral features (See Section~\ref{sec:Obs Strat}), and can even produce convincing false positives. However, when injecting our planetary models into the data we see that we can recover signals at high S/N for a range of molecules (Figure~\ref{fig:real_vs_inj}). In the absence of molecular detections we can therefore assess the sensitivity of our data using grids of injection-recovery tests to constrain the atmospheric properties of the planet (Section~\ref{sec:Inj Tests}).

\subsection{Injection tests}
\label{sec:Inj Tests}

In the event of a non-detection of atmospheric features with HRCCS, injection tests have been widely used to constrain the atmospheric properties of the observed planet (e.g. \citealt{Hoeijmakers2018,Merritt2020,Deibert2021,Spring2022,Lafarga2023,Dash2024,Grasser2024}). We carry out injection tests for a grid of parameters for each molecule we aim to detect, with metallicities ranging from 1 to 1000 times solar metallicity (Z = 1, 10, 100, 150, 200, 250, 500, 1000 Z$_{\odot}$), and cloud top pressures spanning six orders of magnitude from 10$^{1}$ to 10$^{-5}$ bar (log(P) = 1, 0, -1, -2, -3, -4, -5 bar). The contribution functions of the planetary spectra (parametrised by the log opacity Jacobian; e.g. \citealt{August2023}) show sensitivity to atmospheric regions from $\sim$1--10$^{-6}$ bar and thus the highest pressure cloud top tested, 10~bar, represents an effectively clear atmosphere scenario (see Figure \ref{fig:jacobian}). 

It is vital that the injected signals are subjected to identical processing as the real planetary signals and we inject model transmission spectra into the data immediately following spectral extraction and telluric wavelength calibration, but crucially prior to the removal of systematics. However, the impact of correlated noise and line distortions on the model spectra from instrumental effects cannot be accounted for, and therefore the injection tests will always be more sensitive to injected signals than planetary spectral features. Accordingly, we adopt strict 5$\sigma$ upper limits when quoting constraints on planetary properties from our analysis.

Initially, we inject an atmospheric model containing a single molecular species, then cross-correlate with that same model. These simple injections demonstrate that HRCCS with CRIRES+ is theoretically sensitive not only to key molecules such as CH$_4$ and H$_2$O in sub-Neptune atmospheres, but also to trace molecular species including NH$_3$ and H$_2$S (e.g. Figure~\ref{fig:real_vs_inj}). The atmospheric constraints produced from the single-species injection-recovery method are shown in Figure~\ref{fig:injections_ind_inj}. However, this method only provides a measure of the sensitivity of the data to a specific trace species in isolation. While this like-for-like approach may be appropriate for atmospheres dominated by a single trace species \citep[e.g.][]{Lafarga2023, Dash2024, Grasser2024}, it is not necessarily accurate for more complex spectra where a multitude of molecular species can obscure the spectral lines from any individual molecule. This effect is particularly pronounced in the \textit{K}-band, and injections of a single species model are at risk of overestimating the constraining power of the data. We therefore proceed to inject the data with a model containing opacities from \textit{all} our considered species (i.e. CH$_4$, H$_2$O, NH$_3$, H$_2$S, CO$_2$, and CO) at their equilibrium abundances, but then cross-correlate with model templates containing only a single species. This process more faithfully reproduces the analysis procedure used in traditional HRCCS to detect genuine planetary signals, in which a model template containing spectral features from a single species is cross-correlated with the complex spectrum of the observed planet. Consequently, this all-species injection-recovery approach is more robust against the overestimation of atmospheric constraints, though it does inherently assume the choice of the injected all-species model is a good representation of the atmosphere of GJ~3090~b.

\begin{figure*}
    \includegraphics[width=2\columnwidth]{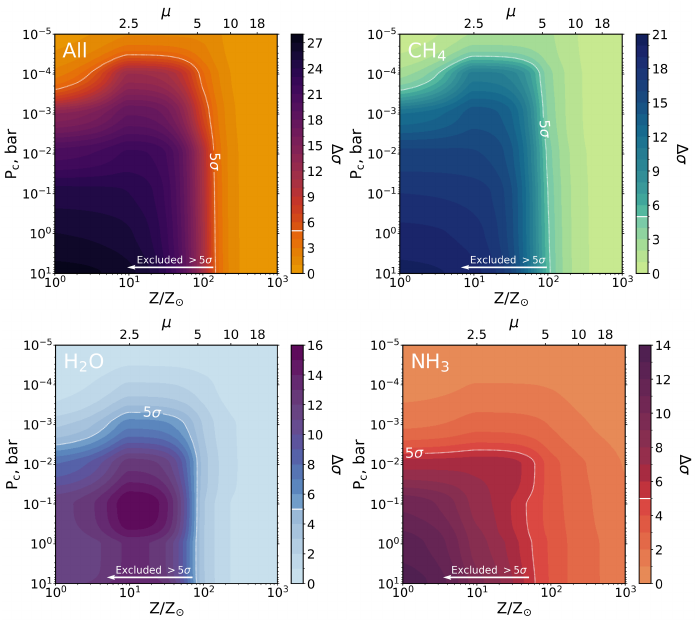}
    \caption{Recovered significance on the planetary signal ($\Delta\sigma$) when performing all-species injection-recovery tests i.e. where a model containing spectral features from all considered molecules (CH$_4$, H$_2$O, NH$_3$, H$_2$S, CO$_2$, CO) is injected, and a template containing only a single molecule is used to recover it. The exception is the top left panel where the cross-correlation template matches the all-species injected model. Dark regions denote areas of the parameter space where the injected model is confidently recovered to $>$5$\sigma$ and thus ruled out as a plausible scenarios for the planetary atmosphere. The sensitivity of the data to CH$_4$ is largely invariant across the all vs. single-species injection methods (see Figure~\ref{fig:injections_ind_inj}). Low metallicities with low cloud decks are strongly disfavoured. The molecules for which we have no constraining power in these injection-recovery tests are shown in Figure~\ref{fig:injections_no_constrainign_power}.
}
    \label{fig:injections_all_inj}
\end{figure*}

The results of these all-species injection-recovery tests are shown in Figure~\ref{fig:injections_all_inj}. These injection tests demonstrate that the data are highly sensitive to a large range of atmospheric scenarios, across multiple molecular species, and demonstrate two overarching trends across the recovered models. First, increasing the atmospheric metallicity (Z), corresponding to an increase in the mean molecular weight (MMW, $\mu$) of the atmosphere through the enrichment of heavier elements, drives a reduction in the atmospheric scale height. This mutes spectral features and results in poor detectability at high metallicity. For H$_2$O the largest spectral features are achieved for a 10$\times$ solar metallicity atmosphere (10~Z$_{\odot}$), as the increased abundances of the targeted molecular species, due to an enriched reservoir of oxygen in the atmosphere, outweighs the impact of the reduced scale height. All species show a sharp drop off in detectability for very high metallicity atmospheric scenarios $\gg$~100~Z$_{\odot}$.

Second, the detectability of a model is reduced by cloud decks at increasingly high altitudes, due to the truncation of spectral lines in the transmission spectrum (Figure~\ref{fig:cloud_models}). While the damping effect of high altitude aerosol layers is dominant at lower spectral resolution, at the resolving power of R~=~146\,000 we see a much weaker dependence of detection significance on cloud top pressure, confirming predictions for observations of cloudy sub-Neptunes with HRCCS \citep{Gandhi2020,Hood2020}. Counter intuitively, the models which can be excluded at highest significance for H$_2$O are at cloud deck pressures of 10$^{-1}$ bar rather than at 10$^{1}$ bar, the lowest altitude cloud pressure in the grid. We suggest that this behaviour arises as the H$_2$O features in the central CRIRES+ K2166 orders (orders 3$\&$4), which additionally contain strong CH$_4$ spectral features, are suppressed for models with higher cloud decks, preventing obscuration of the lines in the H$_2$O model template by the CH$_4$ lines that dominate the injected model, and thus increasing the retrieved significance of the H$_2$O signal. 

The injection tests also illustrate that the data is largely insensitive to the carbon bearing species CO and CO$_2$, even for low metallicity atmospheric scenarios with low altitude clouds (Figure~\ref{fig:injections_no_constrainign_power}). For CO$_2$, this is largely due to the CO$_2$ features in the \textit{K}-band falling in regions of contamination from telluric CO$_2$. The lack of sensitivity to CO is impacted by the strong dependence of the CO lines on the cloud deck pressure and metallicity, and is complicated by strong stellar residuals from deep CO lines in the M-dwarf stellar spectrum, and the impact of the Rossiter-Mclaughlin effect (See Section~\ref{sec:RM_effect}). Uniquely, however, the CO$_2$ injections have marginal sensitivity to very high metallicity scenarios with Z~>~500~Z$_{\odot}$. This is because a higher MMW drives an increasing CO$_2$ abundance due to the enrichment of volatile carbon and oxygen, and thus produces larger spectral features in transmission, despite the decreasing scale height with metallicity. The CO$_2$ constraints from the single species model injections therefore weakly disfavour the highest metallicity scenarios for low altitude clouds at $\sim$3$\sigma$.

The two atmospheric scenarios constrained by this data are degenerate. Using the injected model containing all planetary features, the injection-recovery tests indicate that GJ~3090~b has either a high metallicity (Z~$\gtrsim$~150~Z$_{\odot}$), high mean molecular weight ($\mu$~>~7.1) atmosphere (with an aerosol layer at pressures $\lesssim$10$^{-2}$ bar), or an aerosol layer at pressures <~3.3$\times$10$^{-5}$ bar for which the metallicity is unconstrained.

\subsection{Inclusion of archival CRIRES+ data}
\label{sec:archival_data}

When carrying out the all-species injection-recovery using the full equilibrium chemistry models (Section~\ref{sec:Inj Tests}), the archival K2148 transits are only sensitive to CH$_4$ and H$_2$O. The constraints placed on the atmospheric properties from these two transits alone are limited to the relatively modest 5$\sigma$ constraints on the atmospheric metallicity, Z~>~30 Z$_{\odot}$, and mean molecular weight $\mu$~>~3.9 for cloud decks >~6$\times$10$^{-3}$~bar, and can only exclude cloud decks at a maximum pressure of >~3.3$\times$10$^{-3}$~bar, two orders of magnitude less constraining than the four K2166 transit constraints (See Figure~\ref{fig:injections_K2148}).

The addition of these two archival K2148 transits to the K2166 data set does not therefore offer any significant improvements on the atmospheric constraints. While their inclusion marginally increases the significance at which the most readily detectable (i.e. low metallicity, low altitude cloud deck) scenarios can be ruled out, it also provides a net decrease in the upper limits placed, due to the inclusion of additional noise in the CCFs for the more challenging atmospheric scenarios, for which these archival transits have no constraining power. 

Under idealised photon-dominated Poisson statistics the constraints from 2 transits have 71$\%$ of the sensitivity of 4 transits. However, the disparity in constraining power between the K2166 transits and the archival K2148 transits exceeds this, and is driven primarily by the difference in data quality, as a result of contrasting observing conditions. While the four transits in the K2166 grating are performed under excellent stable seeing conditions (Avg. FWHM of ~0.5\arcsec), the two archival K2148 transits are observed in variable seeing conditions (see Table \ref{tab:obs_table_archival}), with an average FWHM of 0.81\arcsec and 1.10\arcsec, respectively, impacting the stability and S/N of the spectra. The airmass at mid transit for the second archival K2148 transit additionally suffers from higher airmass pre-transit frames, reaching up to an airmass of 1.8. The S/N per exposure is consequently reduced, from an average of 155 for the four K2166 transits, to an average of 89 for the two archival transits, equivalent to a factor of 3 discrepancy in observing time under Poisson statistics.

Furthermore, due to the lower quality and variable seeing conditions, the MACAO system did not consistently suppress the stellar FWHM <~0.2\arcsec, the CRIRES+ slit width, during the archival K2148 transits of GJ~3090~b, resulting in a spectral resolution of R~$\approx$~92\,000. In comparison, the first four transits of GJ~3090~b achieve FWHM <~0.2\arcsec, leading to super-resolution effects and spectral resolutions in excess of R~=~140\,000 (See Section~\ref{sec:Methods}). As the S/N scaling for HRCCS increases with $\sqrt{N_{\text{lines}}}$ lines, this results in an additional theoretical increase in the S/N of 26$\%$ when comparing observations taken with a spectral resolution of R = 92\,000 compared to R = 146\,000, due to the increase in the number of resolvable planetary spectral lines (when using the CH$_4$ models used in this work). A high spectral resolution is particularly valuable for HRCCS observations of hazy sub-Neptunes in transmission, as it allows for the cores of narrow spectral lines truncated by high altitude aerosol layers to be resolved. However, we caution that while the super-resolution is advantageous in this work, it introduces challenges to the data analysis, namely systematic differences between A and B nods and the potential for time varying systematics driven by slit losses if the narrow PSF is permitted to wander in the slit. If the increased spectral resolution exceeds the Nyquist sampling frequency of the detector in the dispersion direction, then the increased spectral resolution will offer diminishing returns due to undersampling of the extracted spectra.

Finally, the two K2148 observations additionally suffer from high and variable precipitable water vapour (PWV) across the transits, reaching maxima of 2.8 ppm and 4.31 ppm. In the near-infrared H$_2$O is a key telluric molecule and the PWV impacts the strength of telluric contamination, while any variability imprints residual noise into the data following PCA-based cleaning \citep{Chiavassa2019,Smith2024}.  Using the simple S/N scaling relation from \citealt{Birkby2018}, we estimate that each individual transit of the four K2166 transits has approximately 2.2 times the sensitivity to the spectral features of GJ~3090~b than each of the archival K2148 transits.

\section{Discussion}
\label{sec:Discussion}

\subsection{The atmosphere of GJ~3090~b within the context of the sub-Neptune population}
\label{sec:population context}

Henceforth, we adopt the constraints from the injection and recovery of the model containing all tested molecules (`All'; Figure~\ref{fig:injections_all_inj}), which imply 5$\sigma$ limits on the metallicity and MMW of Z~>~150 Z$_{\odot}$ and $\mu$~>~7.1 (for a cloud deck $\lesssim$~10$^{-2}$~bar), or an aerosol layer at pressures <~3.3$\times$10$^{-5}$ bar with the metallicity unconstrained. These results have similar constraining power on the atmospheric metallicity to previous studies of the warm-Neptunes GJ~436~b and GJ~3470~b using HRCCS \citep{Grasser2024, Dash2024}, but place constraints on the aerosol layers that are over an order of magnitude tighter. This is remarkable given that GJ~3090~b has approximately half the planetary radius of GJ~436~b and GJ~3470~b. While the scale height of GJ~3090~b assuming a H/He dominated atmosphere (H$_0$ = 340 km) is more favourable than GJ~3470~b (H$_0$ = 335 km) and GJ~436~b (H$_0$ = 210 km), when comparing the transmission spectroscopy metrics GJ~3090~b (TSM = 221) is a significantly more challenging target than GJ~436~b (TSM~=~621) and GJ~3470~b (TSM~=~354). These constraints are therefore testament to both the high quality data obtained with CRIRES+ and the wavelength coverage in the \textit{K}-band, specifically the coverage of CH$_4$ spectral features, which provide the strongest constraints on cloud deck pressure. 

A direct comparison of these CRIRES+ constraints to \textit{JWST} results for GJ~3090~b is not yet possible (See Section \ref{sec:JWST}). However, the sub-Neptunes TOI-836~c \citep{Wallack2024} and TOI-776~c \citep{Teske2025}, both observed in transmission by \textit{JWST}, provide an interesting comparison. The constraints on MMW and cloud deck pressure for GJ~3090~b placed by CRIRES+ are comparable to those placed on TOI-836~c and TOI-776~c, which required 6.8 and 14.6 hours of \textit{JWST} observing time, respectively. Both of these sub-Neptunes have similar radii to GJ~3090~b, but are more challenging targets for transmission spectroscopy due to higher masses and consequently reduced scale heights (TOI-836~c: H$_0$ = 176 km, TOI-776~c: H$_0$ = 97 km, under the assumption of a H/He dominated atmosphere), and thus have reduced transmission spectroscopy metrics (TOI-836~c: TSM~=~104, TOI-776~c: TSM~=~44). Nonetheless, the ability of CRIRES+ observations to place constraints on GJ~3090~b that are comparable to those that \textit{JWST} is achieving for sub-Neptunes (albeit for more challenging targets) demonstrates the potential of HRCCS for the characterisation of sub-Neptunes. 

Furthermore, while we adopt the 5$\sigma$ constraint in our analysis for robustness, we note that our 3$\sigma$ constraints on cloud deck pressure on GJ~3090~b is similar to the 3$\sigma$ constraint of 10$^{-5}$~bar for low metallicity atmospheres on GJ~1214~b (TSM~=~416), from 77 hours of HST observations \citep{Kreidberg2014}. These CRIRES+ observations therefore demonstrate the significant constraining power of ground-based HRCCS on high altitude aerosol layers on hazy sub-Neptunes, requiring approximately one eighth of the observing time compared to HST to place equivalent constraints for a more challenging target.

\begin{figure}
    \includegraphics[width=\columnwidth]{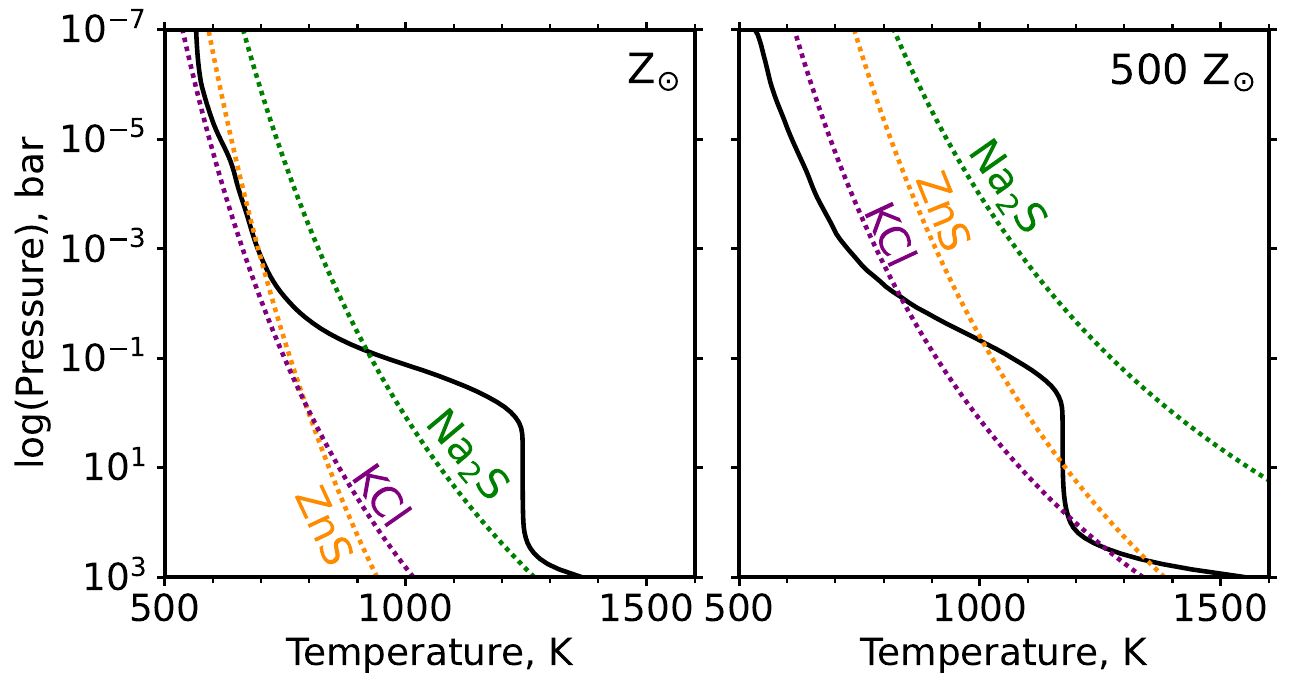}
    \caption{Model temperature-pressure profiles for a solar metallicity (Z~=~Z$_{\odot}$; left) and highly metal enriched atmosphere (Z~=~500 Z$_{\odot}$; right) for GJ~3090~b, produced under radiative-convective equilibrium and for equilibrium chemistry using HELIOS and Fastchem \citep{Malik2017, Malik2019, Stock2018}, see Section~\ref{sec:atmo models}. Condensation curves for key condensible species are plotted for each metallicity case, using the prescriptions from \citet{Morley2012} and \citet{Visscher2006}. For solar metallicity atmospheres ZnS condensates are predicted to be able to form at pressures $\sim$10$^{-4}$~bar, sufficient to match the observed constraints. }
    \label{fig:TP_plot}
\end{figure}

However, the interpretation of the CRIRES+ constraints is hampered by the inherent degeneracy between high altitude aerosols truncating planetary spectral lines in transmission, and thus reducing the detectability of species; and increasing atmospheric metallicity, which shrinks the spectral features. One scenario consistent with these results is that the muted spectral features are driven by high altitude aerosols in a low metallicity (Z~<~100~Z$_\odot$) atmosphere. Radiative-convective models of the vertical temperature structure of GJ~3090~b predict the formation of KCl, Na$_2$S, and ZnS condensates in equilibrium chemistry across a range of atmospheric metallicities (Figure~\ref{fig:TP_plot}), with KCl and ZnS predicted to provide the key contributions to high altitude opacity. 

The well studied warm sub-Neptune GJ~1214~b, observed to host both a high altitude haze layer and a metal rich atmosphere \citep{Kreidberg2014,Kempton2023, Gao2023,Schlawin2024,Ohno2024,Lavvas2024}, and with an equilibrium temperature T$_{\text{eq}}$ = 596$\pm$19~K, provides a compelling comparison to GJ~3090~b (T$_{\text{eq}}$~=~693$\pm$18~K), despite its $\sim$2 times greater mass \citep[8.17~$\pm$~0.43~M$_{\oplus}$;][]{Cloutier2021}. Indeed the predicted condensates for GJ~3090~b mirror the predictions of KCl and ZnS condensates for GJ~1214~b \citep{Morley2013,Kreidberg2014}. For solar metallicity atmospheres for GJ~3090~b, ZnS condensates are predicted to be able to form at pressures $\sim$10$^{-4}$~bar, sufficient to match the observed CRIRES+ constraints at solar metallicity. For atmospheres with increasing mean molecular weights the upper atmosphere cools, and the condensation curves for KCl, Na$_2$S, and ZnS shift to higher temperatures, forming KCl and ZnS clouds at pressures $\sim$10$^{-2}$ bar. To be the source of the non-detection, by causing truncated spectral features at high resolution, these condensates would therefore have to be transported to higher altitude regions of the atmosphere of GJ~3090~b, with pressures <~10$^{-4}$ bar. It has been suggested for GJ~1214~b that either vigorous internal mixing could loft condensates to higher altitudes \citep{Charnay2015b,Charnay2015a}, or that mineral clouds of KCl and ZnS can rise to high altitudes if the cloud particles have sufficiently small radii, and thus have a low settling velocity \citep{Morley2013,Morley2015}. However, theoretical models have struggled to form sufficiently thick clouds at the required altitudes to match the flat transmission spectra of GJ~1214~b \citep{Ohno2018,Adams2019}. 

An alternative scenario for aerosols on GJ 3090 b, proposed to explain the heavily dampened spectral features on GJ~1214~b, suggests a thick photochemical haze providing a strong opacity at NIR wavelengths \citep{Kreidberg2014,Kempton2023}. A wide range of species have been proposed as potential outcomes of haze formation in warm sub-Neptune atmospheres including tholins, hydrocarbon chains analogous to the haze on Saturn’s moon Titan; complex molecules with chemical formulas C$_w$H$_x$N$_y$O$_z$; and diamonds formed through chemical vapour deposition \citep{Horst2018, Gavilan2018,Moran2020,Ohno2024_diamond}. The presence of a high altitude photochemical haze on GJ~3090~b would conform to and corroborate the growing evidence for the trend of haze production with temperature across the sub-Neptune population \citep{Crossfield2017, Brande2024}. Warm sub-Neptunes with 500~$< T_{\text{eq}} <$~800~K have been identified to suffer from the greatest attenuation of spectral features, and GJ~3090~b, with an equilibrium temperature of 693$\pm$18~K, falls within this region of maximum attenuation. 

However, the constraints placed in this work are also consistent with GJ~3090~b hosting a highly metal enriched envelope, with mean molecular weight $\mu$~>~7.1~g~mol$^{-1}$. A high metallicity scenario would be driven by a significant enrichment of CNOPS elements, forming increased molecular abundances of e.g. H$_2$O, CH$_4$, NH$_3$, and CO$_2$ and resulting in a declining mass fraction of H/He \citep{Nixon2024}. Addressing only the constraints on metallicity, the CRIRES+ observations appear consistent with a high metallicity envelope composed of miscible H/He and metals \citep{Benneke2024}. Under the assumption of equilibrium chemistry and a solar C/O~=~0.55, the CRIRES+ observations imply a metal mass fraction Z$_{\text{atm}}$~>~67$\%$ for the low altitude cloud deck scenarios. This in turn provides an upper bound on the H/He envelope mass fraction of $x_{\text{H/He}} < 33 \%$, implying that GJ 3090 b is not H/He dominated if a high metallicity atmosphere drives the non-detections in this work. This estimate relies on the additional assumptions of a well mixed envelope in chemical equilibrium, with a He mass fraction Y~=~0.275. 

While this result is dependent upon the assumption of equilibrium chemistry and the adopted C/O ratio, it is largely insensitive to the exact chemical mixing ratios of the volatile species. These observations therefore disfavour a `Gas Dwarf' composition for GJ~3090~b \citep{Fortney2013, Lopez2014}, in which the atmosphere is dominated by H/He, and has a low mean molecular weight <~3.3 \citep{Benneke2024}. In order for this scenario to exist on GJ~3090~b either photochemical haze or clouds are required to form at pressures <~10$^{-4}$~bar. The high metal mass fraction Z$_{\text{atm}}$~>~67$\%$ is consistent with a `Metal-Rich Miscible Envelope sub-Neptune' proposed by \citep{Benneke2024}, but also approaches the 75$\%$ metal mass fraction proposed to mark the transition to `high-volatile-metal-mass-fraction worlds'. These planets are typified by the `steam world' GJ 9827 d, seen to host spectral features from H$_2$O and a measured MMW of 18~g~mol$^{-1}$ \citep{Piaulet-Ghorayeb2024}, but could theoretically be dominated by any volatile molecular species.

While the degeneracy with high altitude aerosol layers prevents detailed conclusion on the exact composition of the atmosphere of GJ~3090~b, the ability to exclude $\mu$~<~7.1 scenarios for cloud decks >~10$^{-2}$~bar, is consistent with a highly metal enriched atmospheric composition, and a `water world' scenario for GJ~3090~b cannot be excluded. Recent \textit{JWST} observations of GJ~1214~b have raised the possibility of extremely metal enriched envelopes, Z~$>$~1000~Z$_{\odot}$ alongside high altitude haze \citep{Lavvas2024,Schlawin2024,Ohno2024}, and GJ~3090~b also appears to be consistent with a similar composition. A highly metal enriched atmosphere on GJ~3090~b would likewise be consistent with the empirical mass-metallicity trend shown by the Solar System planets \citep{Wakeford2020}. However, these results are markedly different to the clear, low-metallicity atmosphere of TOI-421~b \citep{Davenport2025}, which orbits a G-type host star. The suppression of haze formation on TOI-421~b compared to GJ~3090~b may arise due to the higher equilibrium temperature (T$_{\text{eq}}$~=~920$\pm$24~K) which results in the depletion of CH$_4$, a key haze precursor molecule. The low MMW atmosphere of TOI-421~b suggests distinct atmospheric evolution pathways for sub-Neptunes around FGK stars compared to M-dwarfs, and further highlights the diversity of atmospheres within the sub-Neptune population.

At present, the degeneracy between high altitude clouds and metal enrichment of the atmosphere on GJ~3090~b prevent detailed conclusions on its atmospheric composition and interior structure. Nonetheless, the growing population of characterised warm sub-Neptunes with similar masses, equilibrium temperatures, and stellar environments, (e.g. GJ~1214~b, GJ~9827~d) provide an initial indication that both metal rich atmospheres and high altitude haze are common within this population.

\subsection{HRCCS strategies for M-dwarf host stars}
\label{sec:Obs Strat}

This study is one of only a few HRCCS analyses of planets transiting M-dwarf host stars to date, alongside \citet{Ridden-Harper2023, Dash2024, Cabot2024,Grasser2024}. The profusion of molecular lines in M-dwarf host stars create unique challenges for HRCCS, so we offer here recommendations for improvements in HRS observations and analysis methods.

\begin{figure*}
    \includegraphics[width=2\columnwidth]{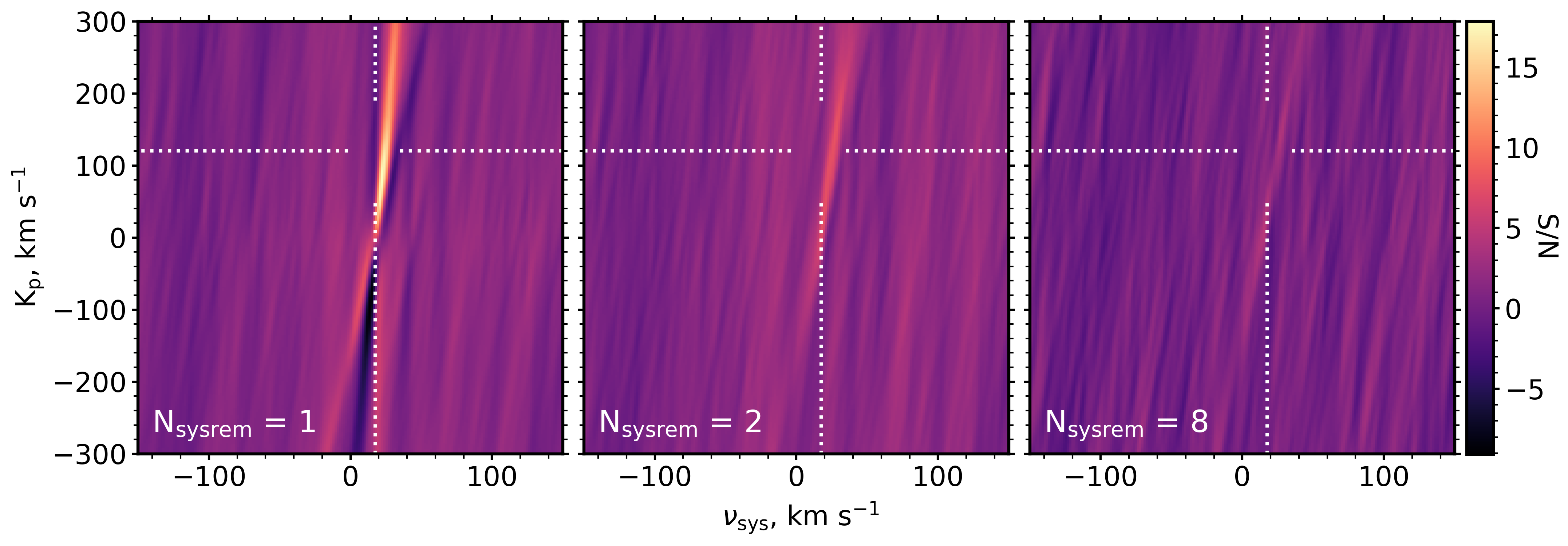}
    \caption{$K_{\text{p}}$--$v_{\text{sys}}$ maps resulting from the cross-correlation of the first night of CRIRES+ spectra with a \textsc{phoenix} stellar model template, demonstrating the evolution of residual M-dwarf features with increasing numbers of $\textsc{sysrem}$ components removed. The stellar residuals dominate following the removal of a single $\textsc{sysrem}$ component (left) with a distinctive asymmetric signal. Following the removal of two $\textsc{sysrem}$ components the stellar residual signal peaks at approximately the planetary $K_{\text{p}}$, masquerading as a redshifted planetary detection with S/N~>~5 (centre). While eight $\textsc{sysrem}$ iterations are sufficient to reduce the stellar residuals to the average noise level of the $K_{\text{p}}$--$v_{\text{sys}}$ map (right), correlated noise still remains at the planetary position. }
    \label{fig:stellar_contamination_plot}
\end{figure*}

\subsubsection{Joint modelling of stellar and telluric contamination}
\label{sec:star_and_telluric_modelling}
The fundamental challenge when observing M-dwarf host stars with HRCCS is the existence of two contaminating signals, in two separate rest-frames, in the form of the spectral features from the star and the tellurics. Uncorrected stellar spectral features induce particularly localised signals in $K_{\text{p}}$--$v_{\text{sys}}$ maps, and can produce false positive signals at the planetary $K_{\text{p}}$ or increase the background noise. We show in Figure \ref{fig:stellar_contamination_plot} the result of cross-correlating the data with an M-dwarf spectrum at various stages of the $\textsc{sysrem}$ detrending. The \textsc{phoenix} stellar models capture the key molecular absorption features from H$_2$O and CO in the M-dwarf photosphere and thus provide a trace of the impact of stellar contamination when performing cross-correlations. However, they are not sufficiently accurate to directly model the observed stellar spectrum.

In this study we mask and detrend the data without separating the blended stellar and telluric lines. However, previous works have successfully modelled and removed stellar or telluric components from the data. For telluric spectral features the spectral synthesis code $\textsc{molecfit}$ \citep{Smette2015} has been successfully applied to high spectral resolution data, principally at visible wavelengths where H$_2$O dominates the telluric spectrum (e.g. \citealt{Allart2017,Hoeijmakers2018_nature, Bello-Arufe2022}). Its use has also been explored for more challenging applications in the near infrared (e.g. \citealt{GonzalezPicos2024}), where the telluric spectrum is denser, and requires the modelling of multiple telluric species (e.g. H$_2$O, CH$_4$, CO$_2$), including saturated telluric lines \citep{Smette2015}. However, the M-dwarf spectrum of GJ 3090 in the \textit{K}-band is dominated by shallow H$_2$O lines across the entire wavelength range, and deep CO lines at the 2.3 $\mu$m band head. These lines blend with features from the tellurics, complicating the fitting of telluric lines with $\textsc{molecfit}$, and additionally remove all remaining continuum, which is required as a baseline by $\textsc{molecfit}$ to provide an accurate fit to the telluric transmission. In order to minimise the contamination from the M-dwarf spectrum, we attempt to use selected regions of the spectra containing relatively isolated and unobscured telluric features to fit a telluric model. However, there are an insufficient number of these unobscured telluric features to capture the behaviour of the multiple telluric molecules that impact our K-band spectra to the required accuracy, and this approach yields inadequate fits to our data.

Previous works have also successfully fit stellar spectral features, although these have been largely limited to narrow spectral regions (e.g. the 2.32~$\mu$m ro-vibrational CO lines in solar-type stars; \citealt{Brogi2016}). The use of numerical three-dimensional hydrodynamical simulations of stellar photospheres has also been explored, and demonstrated to be highly effective at removing the stellar contribution from G and K-type stars \citep{Chiavassa2019, Flowers2019}. However, to date, this approach has not been extended to M-dwarf host stars, and models of stellar photospheres face challenges accounting for the impact of magnetic fields, spot coverage, and activity on synthetic spectra \citep{Allard2013}.

One opportunity for HRCCS is to adapt existing codes commonly used in extreme precision radial velocity (EPRV) planet searches (e.g. \textsc{yarara} \citep{Cretignier2021}; \textsc{wobble} \citep{Bedell2019}; \textsc{soff} \citep{Gilbertson2024}) to perform data driven modelling of the stellar and telluric lines. These methods rely on repeat observations separated by large barycentric velocity shifts to disentangle stellar and telluric signals, suited to EPRV searches. While unfeasible for single transit observations, this analysis method could potentially be applied to multi-transit HRCCS observations, in which the observations are designed such that the epochs in which the transits are observed are well separated in barycentric velocity. Alternative approaches involve the reconstruction of telluric transmission spectra from a previously observed library of telluric standards \citep{Artigau2014_tellcorr}. Existing HRCCS pipelines are implementing these techniques to remove telluric features, including the APERO pipeline for SPIRou/CFHT and NIRPS \citep{Artigau2014Spirou,Pelletier2021,Boucher2023}, and these methods offer a promising approach for removing telluric contamination from HRCCS observations with ELT instrumentation.

\subsubsection{The impact of the Rossiter-Mclaughlin effect}
\label{sec:RM_effect}
Over the course of a planetary transit, the planet occults different regions of the stellar disk, obscuring a fraction of the emitted flux. 
Due to the stellar rotation a net redshift of the integrated stellar line profile will be observed as the planet crosses the approaching blue shifted stellar hemisphere, and correspondingly a net blueshift when the planet crosses the receding redshifted stellar hemisphere. 
The resulting distortion of the stellar spectral line is known as the Rossiter-Mclaughlin (RM) effect or Doppler shadow \citep{Cegla2016,Triaud2018}. 
When using HRCCS to search for a chemical species that is present in both the stellar photosphere and the planetary atmosphere, the distortion from the RM effect will produce signals in the CCF with a non-zero radial velocity shift as a function of time ($K_{\text{RM}}$), and can therefore obscure or mimic a planetary signal \citep{Brogi2016}. As an M2V dwarf, GJ 3090 is expected to contain H$_2$O and CO within its photosphere \citep{Veyette2016}, and we therefore run diagnostic tests to assess the potential impact of the RM effect on our observations.

\begin{figure}
    \includegraphics[width=\columnwidth]{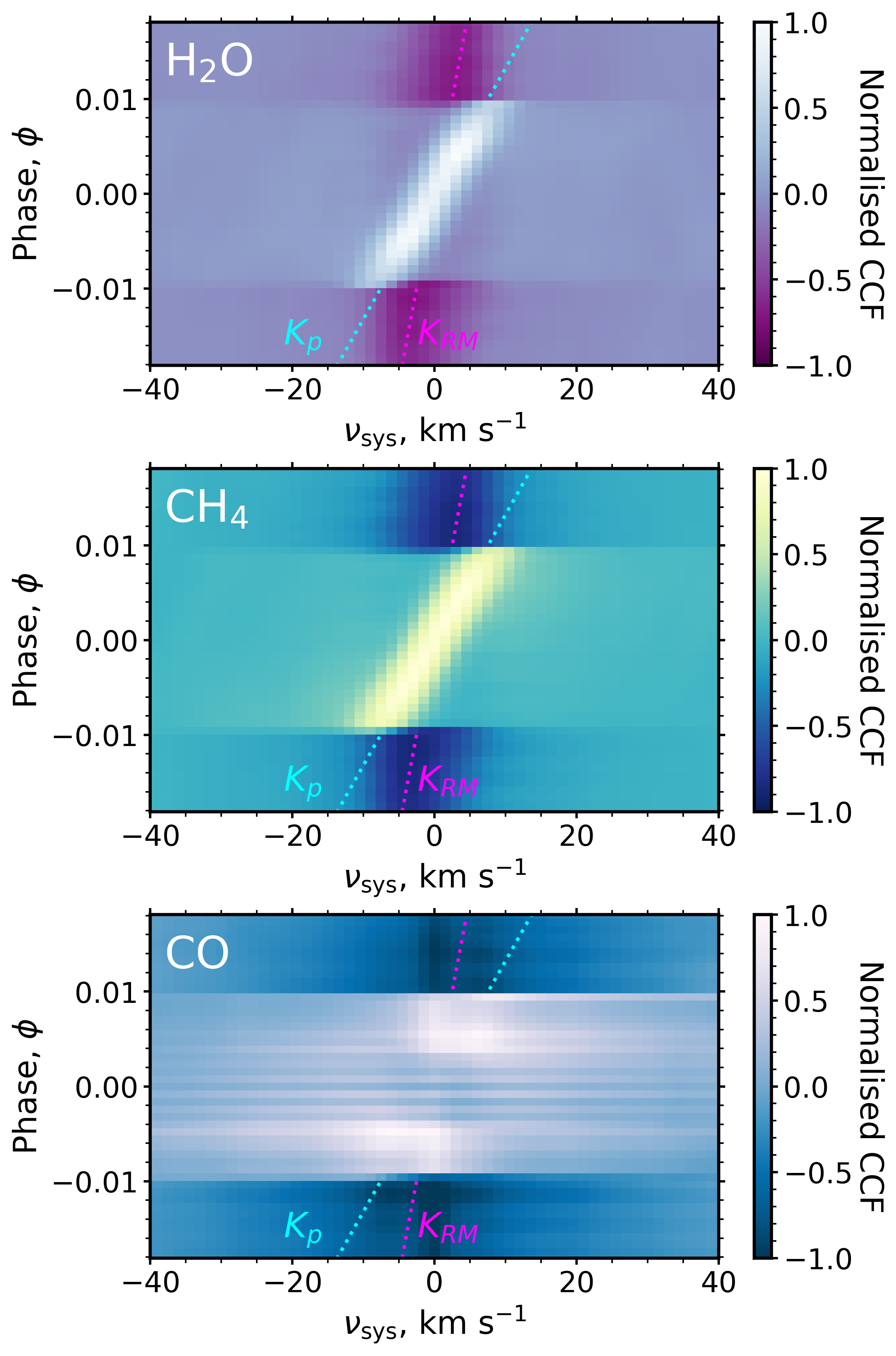}
    \caption{Noiseless simulations of the cross-correlation for GJ 3090 b transiting a rotating stellar disk, processed at the cadence of our observations. Both A and B nods are shown for clarity. Single species models containing H$_2$O, CH$_4$, and CO in a 10 Z$_{\odot}$ metallicity atmosphere with a 10$^1$ bar cloud deck are used for the planetary injection. The injected planetary signals of H$_2$O and CH$_4$ appear largely unaffected by stellar contamination from the Rossiter-Mclaughlin effect, while the CO cross-correlation suffers from severe stellar contamination.}
    \label{fig:RM_plot}
\end{figure}

We forward model a time-series of stellar and planet spectra, considering the stellar rotation and planet orbital motion. The stellar disk is simulated by division into a grid of $500\times500$ cells. Each cell is then assigned a stellar spectrum, for which we use a \textsc{phoenix} model retrieved for the stellar parameters provided in Table~\ref{tab:parameters_table}. The stellar disk is homogeneous, but rotating with $v\sin i = 1.468$~km~s$^{-1}$ \citep{Almenara2022}, in which the contribution from each cell is shifted according to its relative velocity. The planet, with size and coordinates defined by the properties of GJ~3090~b (Table~\ref{tab:parameters_table}) and zero obliquity, is then introduced in transit across the stellar disk. This is initially represented by an opaque core; to include the planetary atmosphere, we take the models described in Section~\ref{sec:CCF models} and resample the opacity from each transited stellar cell according to the $R_\mathrm{p}/R_*$ at each wavelength. We then integrate the stellar and planet contributions, at time-stamps to match those of the observations (Section~\ref{sec:Observations}).

The simulated spectra are subsequently interpolated onto the observed wavelength grid, masked, and divided by the mean spectrum, following the post-processing procedures applied to the real CRIRES+ data, and a single $\textsc{sysrem}$ component is removed. We cross-correlate the residual spectral time-series with the injected model; our results from a single transit are plotted in Figure \ref{fig:RM_plot}. The CH$_4$ CCF shows no in-transit contamination from the RM effect, as is predicted due to the absence of CH$_4$ in the stellar photosphere. H$_2$O, known to be present in the M-dwarf spectra, shows only mild contamination from the Doppler shadow, and the planetary signal is dominant. The RM effect therefore, does not manifestly alter the results of our injection-recovery tests involving H$_2$O, or molecules such as CH$_4$ which are not contained within the stellar photosphere.

Finally, CO shows near-total obscuration of the planetary signal across the transit, with a complex cross-correlation structure composed of three parts; the planetary CO trail, uncorrected stellar residuals, and the RM effect Doppler shadow. The planetary signal is limited to ingress and egress, while interference with the Doppler shadow appears to remove all information contained within the mid-transit frames. Uncorrected stellar residuals, which are constant in phase, remain at ingress and egress, preventing the unique identification of the remaining planetary signal. The introduction of photon noise to the simulated data at the level seen in our observations results in the complete obscuration of the planetary CO features in the CCF. We therefore suggest that the RM effect is partially responsible for the complete lack of sensitivity to CO in the injection-recovery tests. 

Due to the low velocity shift across the transit ($\sim$14 km s$^{-1}$) the Doppler shadow and the planetary signal are coincident in velocity across the transit. Therefore established approaches of masking or modelling the Doppler shadow (e.g. \citealt{Hoeijmakers2020, Prinoth2022, Maguire2023}), risk removing or biasing the entirety of the planetary signal. It is therefore vital to consider the effect of the RM effect on target species when planning observations in the NIR around M-dwarfs, and to simulate the impact at the CCF level to ensure that the detection of atmospheric features is feasible.

\subsubsection{Systemic velocity offsets}
\label{sec:velocity_offsets}
A vital consideration when planning transit HRCCS observations is to ensure that the systemic velocity of the target star relative to the solar barycentre (17.4095~km~s$^{-1}$ for GJ 3090) is misaligned from the barycentric velocity of the Earth at the time of observation \citep{Rodler2014}. For all HRCCS observations this is required to separate the planetary and telluric features in the $K_{\text{p}}$--$v_{\text{sys}}$ parameter space, but the additional inclusion of strong stellar residuals from an M-dwarf host creates a strong false positive risk from overlapping residuals if the systemic and barycentric velocities overlap. All of our nights have a velocity separation >~5~km s$^{-1}$ between the Earth's barycentric velocity offset and the stellar systemic velocity. However, we caution that 5~km~s$^{-1}$ separation is the advisable limit for observations of M-dwarf hosts as the first transit of our dataset, with an offset of 5.7~km~s$^{-1}$, required additional attention while cleaning the data to avoid false positive planetary signals resulting from significant residuals from stellar and telluric overlap.

\subsubsection{High resolution line lists in high MMW atmospheres}
\label{sec:line_lists}
The accuracy of the line lists used in the generation of high-resolution model templates for cross-correlation is crucial for the successful detection of molecular species in exoplanet atmospheres \citep{deRegt2022}. As HRCCS progresses to the study of non-H/He dominated atmospheres, for example the Z~>~100~Z$_{\odot}$ scenarios probed in this work (see Section~\ref{sec:Discussion}), accurate spectral templates are required for the modelling of high mean molecular weight atmospheres. However, 
appropriate line broadening coefficients for non-H/He or non-telluric background gases are not readily available for the majority of atmospheric species of interest \citep{Hedges2016,Barton2017,Vispoel2019, Gharib-Nezhad2019}. With advances in HRCCS and the imminent advent of the ELTs, appropriate atmospheric templates are urgently required to facilitate the reliable and widespread use of HRCCS in high MMW atmospheres.

\subsection{Predictions for \textit{JWST} observations of GJ~3090~b}
\label{sec:JWST}

\begin{figure*}
    \includegraphics[width=2\columnwidth]{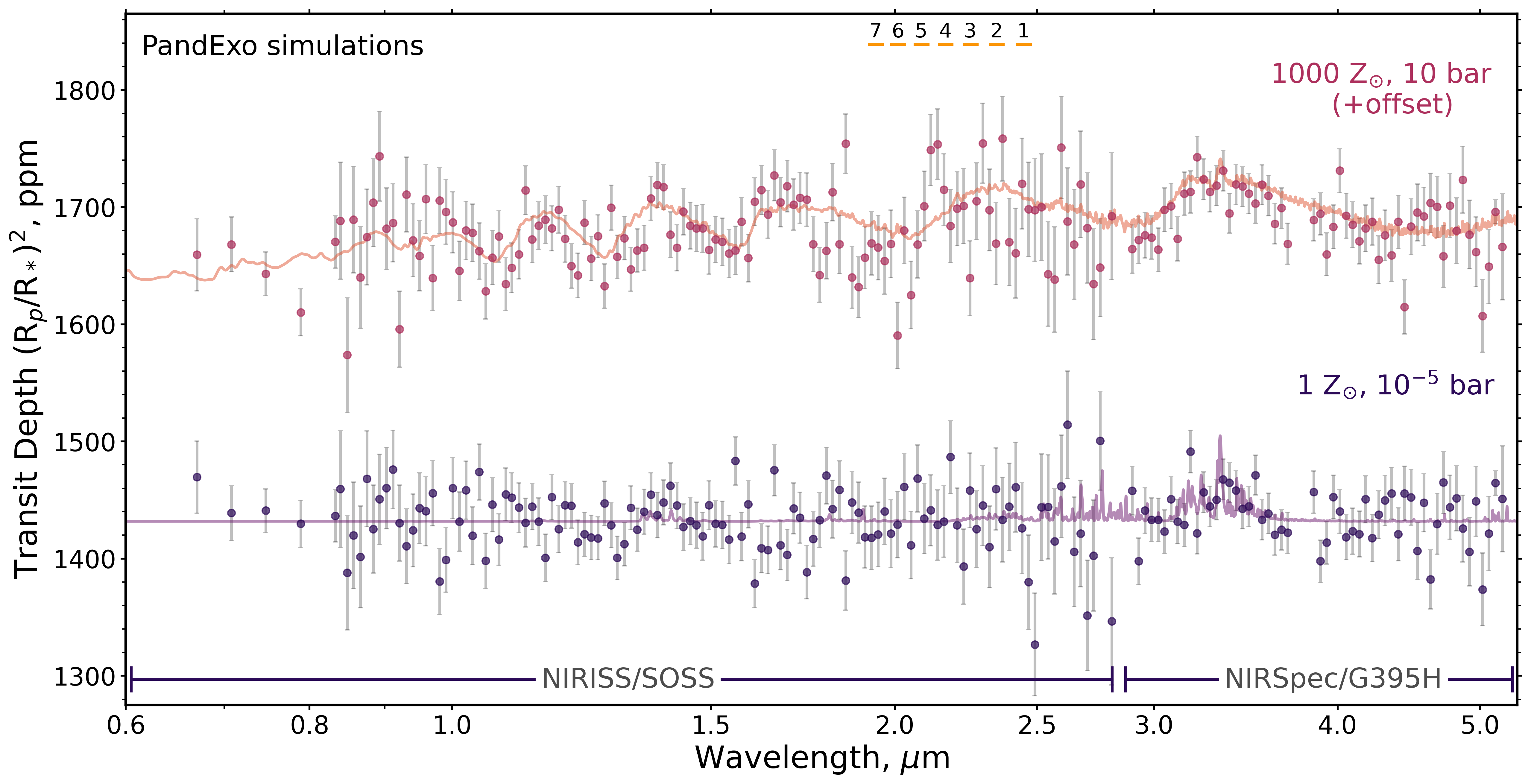}
    \caption{\texttt{PandExo} simulations of NIRSpec G395H (2 transits) and NIRISS/SOSS (2 transits) observations for atmospheric models for GJ~3090~b with a very high metallicity (Z~=~1000~Z$_{\odot}$, 10~bar aerosol layer), and a very high altitude aerosol layer (Z~=~1~Z$_{\odot}$, 10$^{-5}$~bar aerosol layer), following the observing strategy of \textit{JWST} programme ID.4098. The simulated models represent atmospheric scenarios which can not be excluded by our high spectral resolution injection-recovery tests. The simulations do not include potential contamination from the M-dwarf host. \textit{JWST} observations have sensitivity to the atmospheric metallicity through molecular features across the SOSS and G395H wavelength ranges, while high altitude grey aerosol layers obscure the spectral features in the SOSS wavelength range and return heavily muted features in the G395H spectra. The wavelength coverage of the CRIRES+ K2166 orders is denoted in yellow. The G395H and SOSS order 1 spectra are binned to R~=~50, while SOSS order 2 is binned to R~=~10.}
    \label{fig:JWST_plot}
\end{figure*}

GJ~3090~b has been observed in transmission by \textit{JWST}, as part of a programme to observe five potentially volatile-rich sub-Neptunes (ID.4098, PIs: Benneke, Evans-Soma; see \citealt{Benneke2024,Piaulet-Ghorayeb2024}, Ahrer et al. in preparation). Two transit observations were taken with NIRSpec using the G395H setting (R~$\approx$~2700), and a further two transits observed with NIRISS single object slitless spectroscopy (SOSS; R~$\approx$~700), providing spectral coverage from 0.6--5.27~$\mu$m and totalling 20.82 hours of observations. We simulate the detectability of specific planetary models tested in this work with the forthcoming \textit{JWST} observations, using the \textit{JWST} observation simulator \texttt{PandExo} \citep{Batalha2017}. 

Two extreme atmospheric scenarios are presented in Figure~\ref{fig:JWST_plot} that are not ruled out by the CRIRES+ observations. The first is a model with a metallicity of 1000~Z$_{\odot}$ and a low altitude 10 bar cloud deck, and the second an atmosphere with a high altitude aerosol layer at 10$^{-5}$ bar and a solar metallicity. These idealised \textit{JWST} simulations suggest that the upcoming observations have the theoretical sensitivity to detect molecular absorption features in transmission for high metallicity atmospheres, including CH$_4$ and H$_2$O features across the SOSS and G395H spectral ranges. Attenuation of the spectra by high altitude aerosols, under our modelling assumption of a grey cloud deck, has the largest impact in the SOSS spectral range and thus the SOSS data has the largest constraining power on cloud deck pressures at low resolution. The highest altitude aerosol layers tested (<10$^{-5}$ bar) completely obscure spectral features across the entire wavelength range (Figure~\ref{fig:JWST_plot}). For increasingly metal-enriched atmospheres, $>$1000~Z$_{\odot}$, CO$_2$ will dominate the atmospheric opacity, and may provide the most detectable signature of an extremely high metallicity atmosphere. While the 1.92--2.47~$\mu$m CRIRES+ observations presented in this work have limited sensitivity to CO$_2$ due to extreme telluric contamination and can only place $\sim$3$\sigma$ constraints on its presence, CO$_2$ is theoretically detectable in the NIRSpec G395H spectral range at 4.4~$\mu$m. With sufficient S/N from JWST the degeneracy between cloud deck pressure and atmospheric metallicity for GJ 3090 b could therefore be broken. 

However, the impact of stellar contamination must also be considered. At a high spectral resolution the planetary atmosphere lines are resolved from the contaminating M-dwarf lines, due to their relative Doppler separation. At low spectral resolution these lines are blended and the presence of unocculted inhomogeneities (e.g. spots and faculae) in the photosphere of the M-dwarf host can directly contaminate the observed planetary transmission spectrum. This effect, known as the transit light source effect (TLSE; \citealt{Rackham2018}) is wavelength-dependent and can result in spurious spectral signals, potentially orders of magnitude larger than planetary spectral features. While stellar contamination from the TLSE is principally a risk at the 0.6–-2.8 $\mu$m SOSS and 0.6–-5.3 $\mu$m NIRSpec Prism wavelengths \citep{Lim2023,Radica2024, Cadieux2024}, previous \textit{JWST} studies of planets orbiting M-dwarfs have suffered from severe stellar contamination extending to 5~$\mu$m (e.g. \citealt{May2023,Moran2023}), preventing the unambiguous detection of species in the planetary atmospheres. However, we note that stellar contamination from the TLSE is not a universal issue for all M-dwarf hosts (e.g. \citealt{Benneke2024}), and that features from species that are not contained within the stellar photosphere (e.g. CH$_4$, CO$_2$), may still be detectable in the case of stellar contamination (e.g. \citealt{Fournier-Tondreau2024}).

HRCCS observations are highly complementary to low-resolution observations from \textit{JWST}, as they probe distinct atmospheric pressures \citep{Brogi2019}, and joint retrievals of \textit{JWST} and ground-based high-resolution data present a promising avenue to provide tighter constraints on planetary properties \citep{Smith2024}. HRCCS can provide an independent probe of cloud deck pressures to break degeneracies in \textit{JWST} data, as the 0.6--2.8~$\mu$m SOSS wavelengths, which provide the greatest sensitivity to aerosol layers with \textit{JWST}, are at risk of stellar contamination from M-dwarfs. Furthermore, while \textit{JWST} is theoretically sensitive to very high mean molecular weight atmospheres, in the case of a strongly obscured 0.6--2.8~$\mu$m SOSS spectrum (by either aerosol layers or stellar contamination), the MMW estimate can only be inferred from a single CH$_4$ or CO$_2$ feature in the NIRSpec G395H range, introducing considerable uncertainty  \citep{Benneke2013,Benneke2024}. HRCCS can therefore provide highly complementary MMW constraints which are less directly impacted by stellar contamination and high altitude aerosols.

\subsection{Implications for future HRCCS studies of sub-Neptunes}
\label{sec:Future}

GJ~3090~b is the sub-Neptune with the 2\textsuperscript{nd} highest TSM discovered to date (221$^{+66}_{-46}$), and only GJ 1214 b is theoretically more readily observable in transmission. Through injection-recovery tests we have demonstrated that the CRIRES+ data is sensitive to a large range of atmospheric scenarios and thus low metallicity atmospheres with low altitude cloud decks on warm sub-Neptunes orbiting M-dwarfs can theoretically be observed using established HRCCS techniques. However, the non-detections presented in this work suggest that warm sub-Neptunes with high metallicity atmospheres and/or high-altitude aerosol layers remain beyond the observability of present instrumentation using only modest investments of observing time. While we have demonstrated the sensitivity of HRCCS to the cores of spectral lines that extend above high altitude cloud decks (up to 3.3$\times$10$^{-5}$ bar in injection tests), the detection of spectral features from atmospheres with Z $\gg$ 150 Z$_{\odot}$ 
is currently prevented for any injected model by the reduced atmospheric scale height, resulting in muted spectral features. Nonetheless, developments in stellar or telluric modelling, or novel detrending processes may offer some improvements in the sensitivity of HRCCS to these atmospheres (Section~\ref{sec:Obs Strat}).
Alternative avenues using existing HRCCS instrumentation could involve investigating sub-Neptune targets which have been predicted from population studies to have spectra less significantly attenuated by aerosols (e.g. 500 $> T_{\text{eq}} >$ 800 K; \citealt{Crossfield2017,Brande2024}).

The advent of the Extremely Large Telescopes (ELTs), will represent a step-change in the achievable S/N for the observation of exoplanetary atmospheres, nominally a factor of $\sim$5 increase in the S/N on the planetary features based solely on scaling the primary mirror diameter from the VLT, and under the assumption of photon dominated noise statistics. However, we caution that a myriad of instrumental parameters including, but not limited to, the wavelength coverage, relative throughput, instrument stability, achieved spectral resolution, and instrument specific systematic effects will all substantially impact the relative S/N achieved with ELT instrumentation (e.g. ANDES, METIS). Comprehensive end-to-end simulations, including all astrophysical and instrumental effects, are required to simulate HRCCS science cases for ELT instruments (e.g. \citealt{Vaughan2024})

The S/N increase of the ELT alone will not guarantee the success of challenging observations, including those of hazy metal rich sub-Neptunes around M-dwarfs. Alongside the enduring observational challenges of telluric contamination, stellar residuals, and instrumental systematics; stellar effects including the RM effect, the transit light source effect, centre-to-limb variations, and stellar spectral variability over the observations will become increasingly prominent for observations with the sensitivity of the ELTs. These effects, while obscured at the present S/N, likely already inhibit our cleaning processes for GJ~3090~b (e.g. Section~\ref{sec:RM_effect}), and understanding their impact on HRCCS observations prior to observations with the ELT is vital. 

Of particular promise for the study of warm sub-Neptunes with the ELTs are the \textit{M}-band wavelengths (4-5$\mu m$) covered by METIS/ELT \citep{Brandl2021}, and planned for GMTNIRS/GMT \citep{Lee2022}, and MICHI/TMT \citep{Packham2018}. This has been identified as the optimum spectral band for simultaneously detecting multiple species (CO, H$_2$O, CO$_2$, Haze species) in hazy sub-Neptunes in transmission \citep{Hood2020}, despite challenges from telluric contamination and thermal background, and the use of the \textit{M}-band for HRCCS observations has recently been verified with CRIRES+ \citep{Parker2024}.

Moving beyond sub-Neptunes, the detection of molecular features in the atmospheres of rocky exoplanets around M-dwarf hosts using HRCCS in transmission is a key science goal for ELT instrumentation (e.g. ANDES/ELT; \citealt{Palle2023}). This challenging science case represents an order of magnitude increase in the sensitivity required on transit spectral features, from the scale heights to which we are sensitive to for GJ~3090~b ($\gtrsim$100~km), to scale heights of the order of 10~km expected for truly Earth-like exoplanetary atmospheres. The observations of challenging targets requires the stacking of cross-correlation functions from many transits, even with the ELTs \citep{Hardegree-Ullman2023,Currie2023}. Given the risk of inducing false positives when performing this stacking, further work is required to fully explore the impact of combining transits, and to identify diagnostics to delineate real and spurious signals \citep{Cheverall2024}.

\section{Conclusions}
\label{sec:Conclusions}

Using four CRIRES+ \textit{K}-band (K2166) transits of the warm sub-Neptune GJ~3090~b we present the first dedicated search for molecular features in the atmosphere of a short period sub-Neptune using ground-based high-resolution spectroscopy. Despite achieving excellent data quality across all four transits we detect no molecular species. Through injection-recovery tests, we demonstrate the sensitivity of the data to CH$_4$, H$_2$O, NH$_3$ and H$_2$S, and verify the ability of HRCCS to access high aerosol layers on sub-Neptunes, placing constraints similar to those achieved with dedicated space-based observatories (e.g. \citealt{Kreidberg2014}), in a fraction of the observing time. The inclusion of two archival CRIRES+ transits taken in the K2148 grating setting observed in more challenging observing conditions do not improve our constraints, demonstrating the importance of acquiring excellent quality data for this challenging science case.

The injection tests are consistent with two degenerate scenarios for the atmosphere of GJ~3090~b. First, GJ~3090~b may host a highly enriched atmosphere with > 150 Z$_{\odot}$ and mean molecular weight >~7.1 g~mol$^{-1}$, representing a volatile dominated envelope with a H/He mass fraction $x_{\text{H/He}} < 33\%$, with an unconstrained aerosol layer. Second, the data are consistent with a high altitude aerosol layer at pressures < 3.3$\times$10$^{-5}$ bar, with the metallicity unconstrained. Cloud condensates are challenging to form at these altitudes on GJ~3090~b across a range of metallicities, and this attenuation by aerosols may therefore be driven by photochemical haze (Section \ref{sec:Discussion}). Comparing to the growing population of characterised sub-Neptune atmospheres in this temperature range, both metal enrichment and high altitude hazes remain viable scenarios within the warm sub-Neptune population.

Upcoming \textit{JWST} observations of GJ~3090~b (ID.4098, PIs: Benneke, Evans-Soma) have the theoretical sensitivity to detect planetary spectral features in high metallicity atmospheres, and can therefore verify and extend the atmospheric constraints presented in this work. Through the access to high cloud decks HRCCS observations are highly complementary to low-resolution observations from \textit{JWST}, and the combination of the unique sensitivities of the two methods has the potential to place powerful joint constraints on the atmospheric properties of GJ 3090 b.

Finally, we discuss the opportunities and challenges of observing the warm sub-Neptune population with both existing and future instrumentation for the ELTs. The M-dwarf host star poses specific challenges to the HRCCS analysis. Unlocking the full potential of the ELTs for atmospheric characterisation will require significant improvements in the joint modelling and removal of telluric and M-dwarf spectral features, the calculation of molecular broadening coefficients for high mean molecular weight atmospheres, and a precise understanding of the impact of spectral variability from stellar inhomogeneities on high resolution spectra. Pathfinding studies using existing instrumentation to test ELT science cases are a crucial preparatory step to identify challenges prior to first light of the ELT.

\section*{Acknowledgements}

We thank the referee for their helpful comments that improved the quality of the manuscript. We thank Eva-Maria Ahrer and Lisa Nortmann for helpful discussions. We additionally thank ESO astronomers Cedric Ledoux and Michael Abdul-Masih for taking the 109.232F observations and ensuring the excellent data quality which has permitted this work. A special thank you to Ferdinando Patat in the ESO Observing Programmes Office for helping to make this observing program possible.

LTP, JLB, SRV, and CF acknowledge funding from the European Research Council (ERC) under the European Union’s Horizon 2020 research and innovation program under grant agreement No 805445. JLB further acknowledges the support of the Leverhulme Trust via the Philip Leverhulme Physics Prize. JMM acknowledges support from the Horizon Europe Guarantee Fund, grant EP/Z00330X/1. A part of AB-A's contribution to this work was carried out at the Jet Propulsion Laboratory, California Institute of Technology, under a contract with the National Aeronautics and Space Administration (80NM0018D0004).

This study is based on observations collected at the European Organisation for Astronomical Research in the Southern Hemisphere under ESO programme 109.232F. This research has made use of the NASA Exoplanet Archive, which is operated by the California Institute of Technology, under contract with the National Aeronautics and Space Administration under the Exoplanet Exploration Program. This research has made use of NASA’s Astrophysics Data System Bibliographic Services and the SIMBAD database, operated at CDS, Strasbourg, France. This research made use of SAOImageDS9, a tool for data visualization supported by the Chandra X-ray Science Center (CXC) and the High Energy Astrophysics Science Archive Center (HEASARC) with support from the \textit{JWST} Mission office at the Space Telescope Science Institute for 3D visualization \citep{Joye2003}. This work has made use of the Python programming language\footnote{\url{https://www.python.org/}}, in particular packages including NumPy \citep{Harris2020}, SciPy \citep{Virtanen2020}, Matplotlib \citep{Hunter2007}, and Astropy \citep{Astropy2013,Astropy2018,Astropy2022}.

\section*{Data Availability}
The raw data used in this study is available for download from the ESO Data Archive under Programme IDs 109.232F and 0111.C-0106. Processed data products and models are available on reasonable request to the corresponding author.



\bibliographystyle{mnras}
\bibliography{GJ3090}
\FloatBarrier




\appendix
\label{sec:Appendix}

\section{CRIRES+ reduction details}
\label{app:CRIRES+_reduction}
In the raw CRIRES+ K2166 frames we observe systematic vertical stripes in the detectors and assess that these features are likely readout artefacts. The CRIRES+ detectors contain a region at the base of the detector array which is baffled to prevent its exposure to light, and we therefore subtract the median of this region to calibrate out these artefacts. This is set by the \texttt{--subtract$\_$nolight$\_$rows = True} command in the esorex background subtraction and spectral extraction routines. We additionally find that the subtraction of a column-by-column fit to the pixel values between each order aids this calibration, and therefore set \texttt{--subtract$\_$interorder$\_$column = True} during the background subtraction stage. In spectral extraction we find a poor performance of the cosmic ray detection algorithm for this data set, and therefore apply a custom bad pixel detection routine following extraction (see main text, Section~\ref{fig:data_proc}). When extracting individual spectra we set the extraction oversampling to 13, and the swath width to 600 pixels, determined through testing to provide clean extraction and to remove both extraction artefacts on the order of individual pixels seen for lower extraction oversampling, and discontinuities in the continuum observed for lower swath widths. 

\begin{table}
	\centering
	\caption{The number of \textsc{sysrem} components removed for each order (O1-7) and transit. Transits one to four, observed in the K2166 grating setting, form our primary data set while the K2148 transits form the archival data set. \label{tab:sysrem_table}}
	\begin{tabular}{lcccccccc} 
		\hline
             & Grating & O1 & O2 & O3 & O4 & O5 & O6 & O7\\
		\hline
            \hline
            Transit 1 & K2166 & 8 & 9 & 7 & 7 & 7 & 7 & 10\\
            Transit 2 & K2166 & 9 & 9 & 10 & 9 & 7 & 6 & 7\\
            Transit 3 & K2166 & 9 & 8 & 7 & 8 & 8 & 9 & 6\\
            Transit 4 & K2166 & 8 & 9 & 10 & 7 & 7 & 9 & 10 \\
            Transit 5 & K2148 & 10 & 10 & 10 & 8 & 9 & 10 & -\\
            Transit 6 & K2148 & 5 & 8 & 8 & 5 & 8 & 5 & -\\

		\hline
            \hline
	\end{tabular}
\end{table}

\section{Details of the archival K2148 transits}

\begin{table}
	\centering
	\caption{Details of the archival 1.97--2.45~$\mu$m K2148 CRIRES+ observations of GJ~3090~b, denoted as Transit 5 and Transit 6.} 
	\label{tab:obs_table_archival}
	\begin{tabular}{lcc} 
		\hline
             & Transit 5 & Transit 6\\
		\hline
            \hline
            UTC Date & 11/08/2023 & 03/09/2023\\
            Avg.~PWV & 2.14~mm & 4.10~mm\\
            Avg.~Seeing & 0.81\arcsec & 1.10\arcsec \\
            Avg.~Airmass & 1.10 & 1.41 \\
            N$_{\text{exp}}$ & 58 & 52 \\
            DIT & 180~s & 180~s \\
            Slit Width & 0.2\arcsec & 0.2\arcsec \\
            Resolution & $\sim$92\,000 & $\sim$92\,000 \\
            $v_{\text{bary}}$ at T$_0$ & 10.94~km~s$^{-1}$ & 4.48~km~s$^{-1}$ \\
		\hline
            \hline
	\end{tabular}
\end{table}

\begin{figure*}
    \includegraphics[width=2\columnwidth]{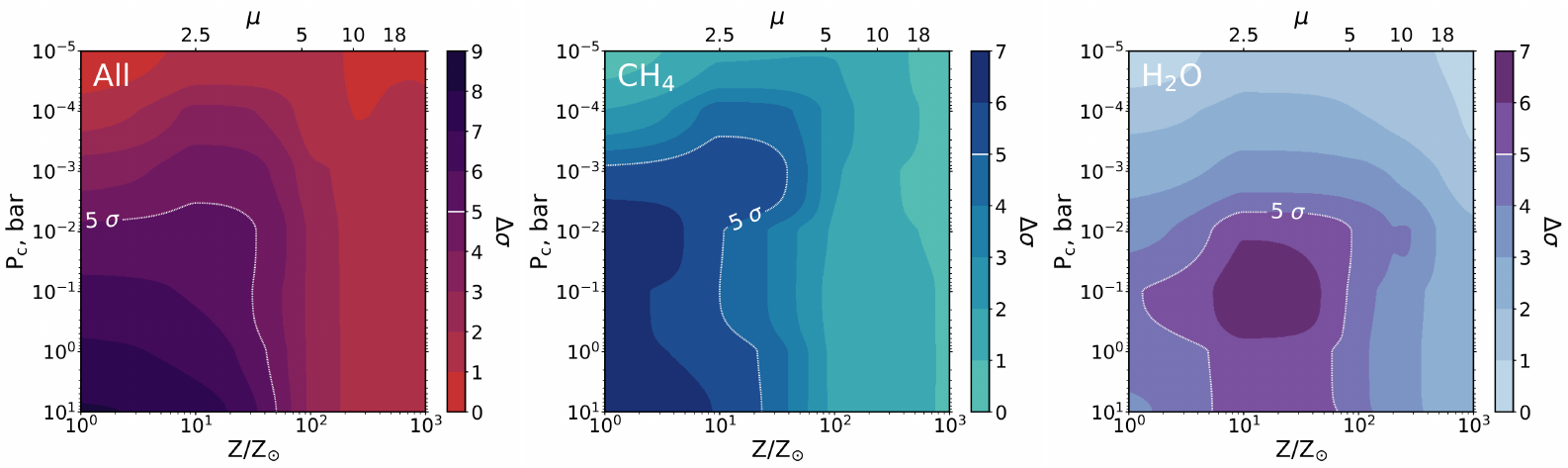}
    \caption{Same as Figure \ref{fig:injections_all_inj} for the all-species injection recovery tests, but for the two archival K2148 transits. The archival transits place 5$\sigma$ constraints on the model with full equilibrium chemistry (labelled All), and for CH$_4$ and H$_2$O, but not for any other considered species. Overall, the archival K2148 transits are two orders of magnitude less constraining on the pressure of the aerosol layer than the constraints from the four K2166 transits.}
    \label{fig:injections_K2148}
\end{figure*}

\section{Additional figures}

\begin{figure}
    \includegraphics[width=\columnwidth]{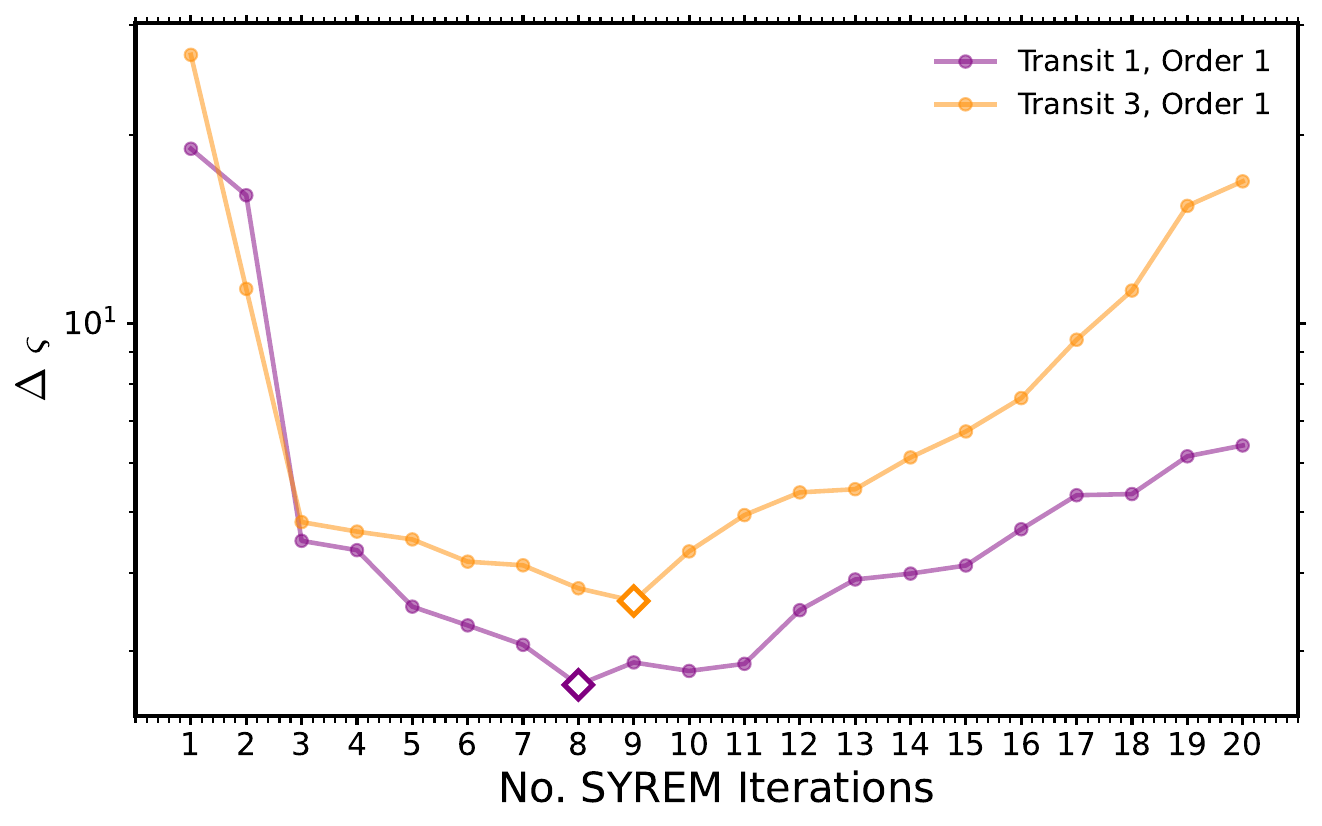}
    \caption{An example of the method used to determine the number of \textsc{sysrem} components to remove from each order following the criteria of Spring $\&$ Birkby (in preparation), see Section~\ref{sec:post_proc}. }
    \label{fig:sysrem_it}
\end{figure}

\begin{figure}
    \includegraphics[width=\columnwidth]{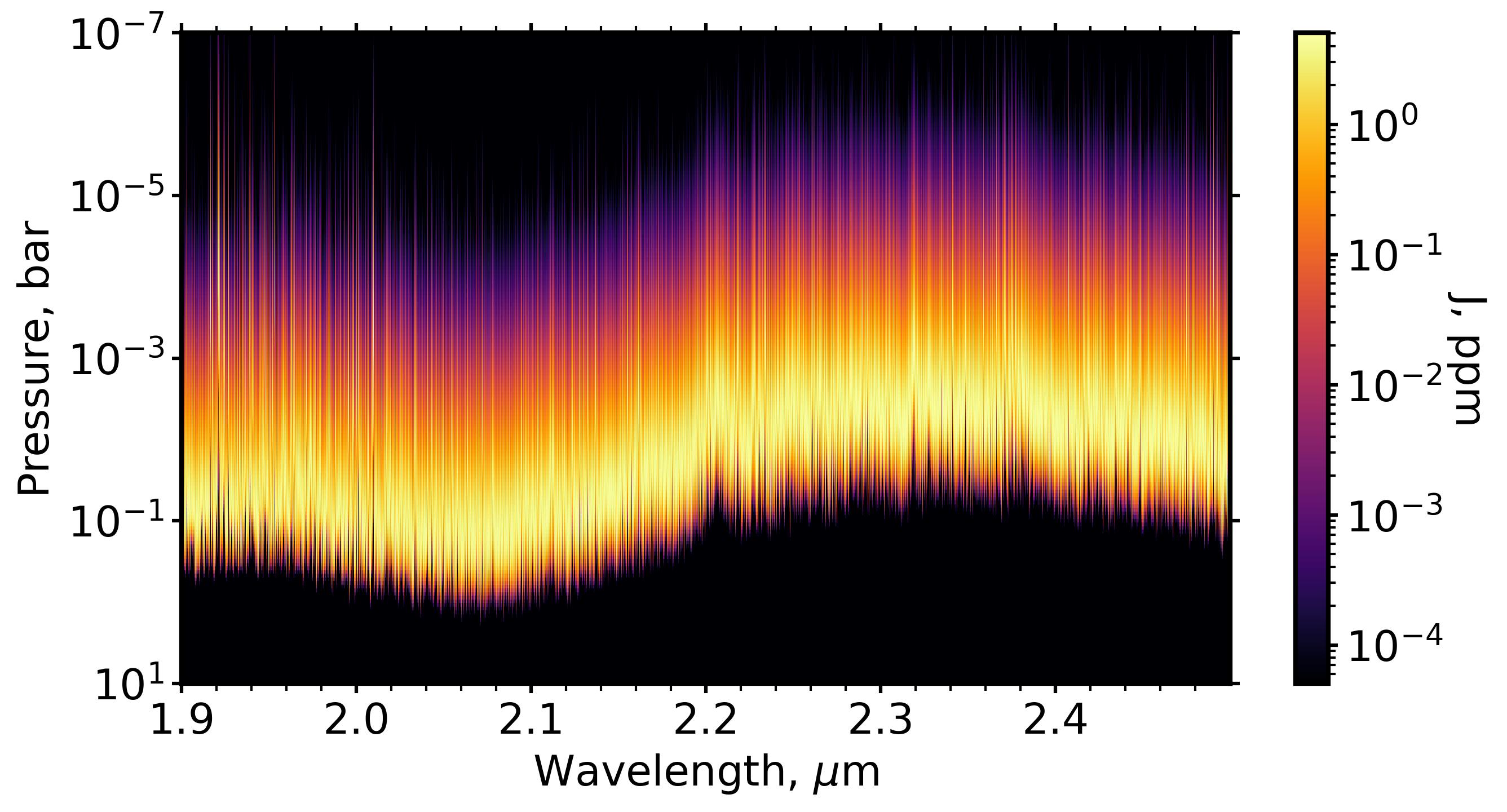}
    \caption{The log opacity Jacobian (J), demonstrating the atmospheric regions of GJ~3090~b probed by the model with solar metallicity, a 10 bar aerosol layer, and equilibrium chemistry.}
    \label{fig:jacobian}
\end{figure}

\begin{figure*}
    \includegraphics[width=2\columnwidth]{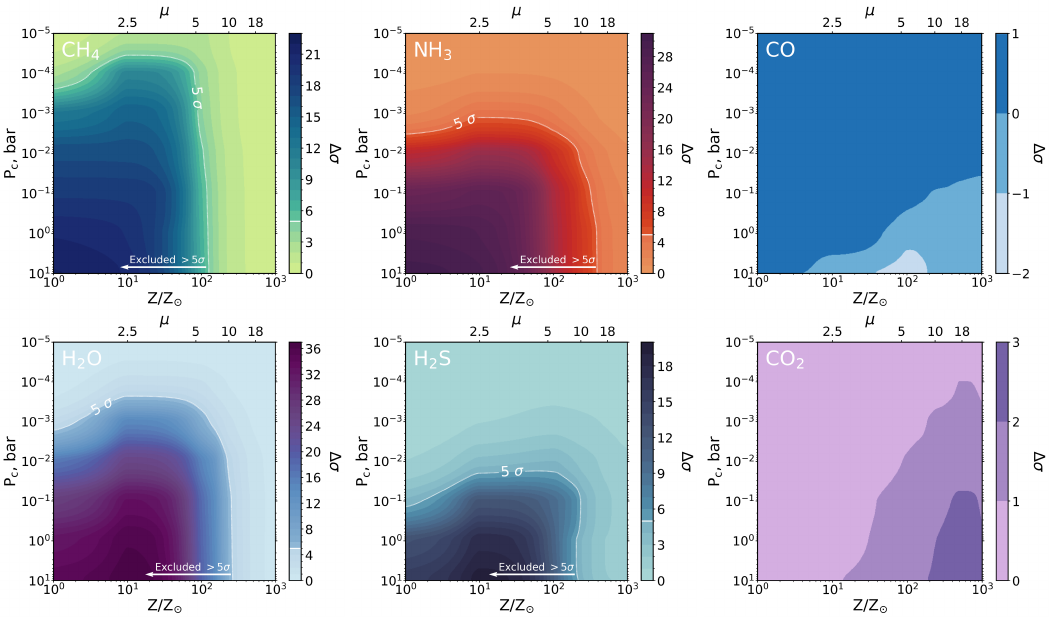}
    \caption{Recovered significance of the planetary signal ($\Delta\sigma$) when performing single-species injection-recovery tests across a grid of metallicities and cloud deck pressures, for CH$_4$, H$_2$O, H$_2$S, NH$_3$, CO, and CO$_2$. Dark regions denote areas of the parameter space where the injected model is confidently recovered to $>$5$\sigma$ and would thus be ruled out as a plausible scenarios for the planetary atmosphere. Negative values of $\Delta \sigma$ denote that the recovered signal at the planet position is disfavoured compared to the measurement of the noise. We caution that single-species injection-recovery tests are at risk of over-constraining the planetary atmosphere due to the absence of obfuscating spectral features from other molecules.}
    \label{fig:injections_ind_inj}
\end{figure*}

\begin{figure*}
    \includegraphics[width=2\columnwidth]{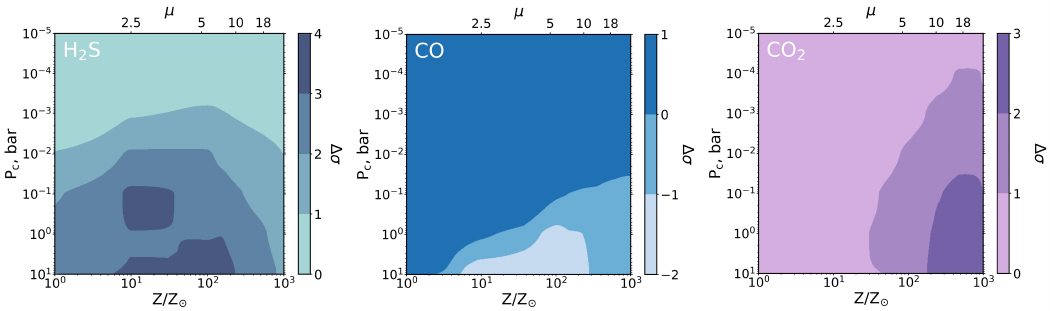}
    \caption{Same as Figure \ref{fig:injections_all_inj} for the all-species injection-recovery tests, showing model templates with no significant constraining power (H$_2$S, CO, and CO$_2$). Negative values of $\Delta \sigma$ denote that the recovered signal at the planet position is disfavoured compared to the measurement of the noise.
    }
    \label{fig:injections_no_constrainign_power}
\end{figure*}


\bsp	
\label{lastpage}
\end{document}